\documentclass[runningheads,a4paper]{llncs}

\usepackage{tikz}
\usepackage{filecontents}
\usepackage{algorithmic}
\usepackage[ruled,vlined]{algorithm2e}

\usepackage{array,multirow}
\usepackage{xspace}
\usepackage{graphicx}
\usepackage{threeparttable}
\usepackage{xspace}
\usepackage{tikz}
\usepackage{enumitem}
\usepackage{url}
\usepackage{xcolor,colortbl}
\definecolor{green}{rgb}{0.1,0.1,0.1}
\definecolor{light-gray}{gray}{0.8}



\usepackage{booktabs}
\usepackage[font=scriptsize,labelfont=bf]{caption,subcaption}

\usepackage{textcomp}
\usepackage{booktabs}
\usepackage[font=small,labelfont=bf]{subcaption}
\usepackage[font=small,labelfont=bf]{caption}
\usepackage{comment}
\usepackage{amsmath,amssymb,amsfonts}

\usepackage{pifont}
\newcommand{\cmark}{\ding{51}}%
\newcommand{\xmark}{\ding{55}}%
\newcommand{\rmark}{\ding{109}}%
\newcommand{\tool}{\texttt{AEOLUS\xspace}}

\newcommand{\vcs}{VDS\xspace}

\begin{document}

\title{Practical Speech Re-use Prevention in Voice-driven Services}

\author{Yangyong Zhang\inst{1}
\and
Maliheh Shirvanian\inst{2}
\and
Sunpreet S. Arora\inst{2}
\and
Jianwei Huang\inst{1}
\and
Guofei Gu\inst{1}
}
\authorrunning{Y. Zhang et al.}
%
\institute{Texas A\&M University, College Station TX 77843, USA\\
\email{\{yangyong, hjw\}@tamu.com}, \email{guofei@cse.tamu.edu}\\
\and
Visa Research, Palo Alto CA 94306, USA\\
\email{\{mshirvan,sunarora\}@visa.com}}
\maketitle              

\begin{abstract}

Voice-driven services (VDS) are being used in a variety of applications ranging from smart home control to payments using digital assistants. The input to such services is often captured via an open voice channel, e.g., using a microphone, in an unsupervised setting. One of the key operational security requirements in such setting is the freshness of the input speech. We present \tool, a security overlay that proactively embeds a dynamic acoustic nonce at the time of user interaction, and detects the presence of the embedded nonce in the recorded speech to ensure freshness. We demonstrate that acoustic nonce can (i) be reliably embedded and retrieved, and (ii) be non-disruptive (and even imperceptible) to a VDS user. Optimal parameters (acoustic nonce's operating frequency, amplitude, and bitrate) are determined for (i) and (ii) from a practical perspective. Experimental results show that \tool\ yields 0.5\% FRR at 0\% FAR for speech re-use prevention upto a distance of 4 meters in three real-world environments with different background noise levels. We also conduct a user study with 120 participants, which shows that the acoustic nonce does not degrade overall user experience for 94.16\% of speech samples, on average, in these environments. \tool\ can therefore be used in practice to prevent speech re-use and ensure the freshness of speech input.

\end{abstract}


\section{Introduction}
\label{sec:intro}

Voice-driven services (VDS) are widely deployed in commercial products to enable personalized and convenient experiences for consumers. Examples include digital assistants from Amazon, Apple and Google for performing tasks like smart home control and online commerce in a hands-free manner, and automated and human-assisted voice response systems for customer support. While this increasing ubiquity can be primarily attributed to improved real world performance of deep learning driven speech and speaker recognition, security is an equally important consideration in operational settings. To secure \vcs in practice, one of the common security mechanisms used is ``voice password''. For example, applications such as Samsung Bixby, Wechat and others~\cite{bixbypassword,wechatpassword,mobileapp_password} prompt users to speak a password and perform two factor authentication by checking (i) if the spoken passphrase is correct, and (ii) it was spoken by the authorized user.

\begin{figure}[t]
\centering
\includegraphics[width=0.8\linewidth]{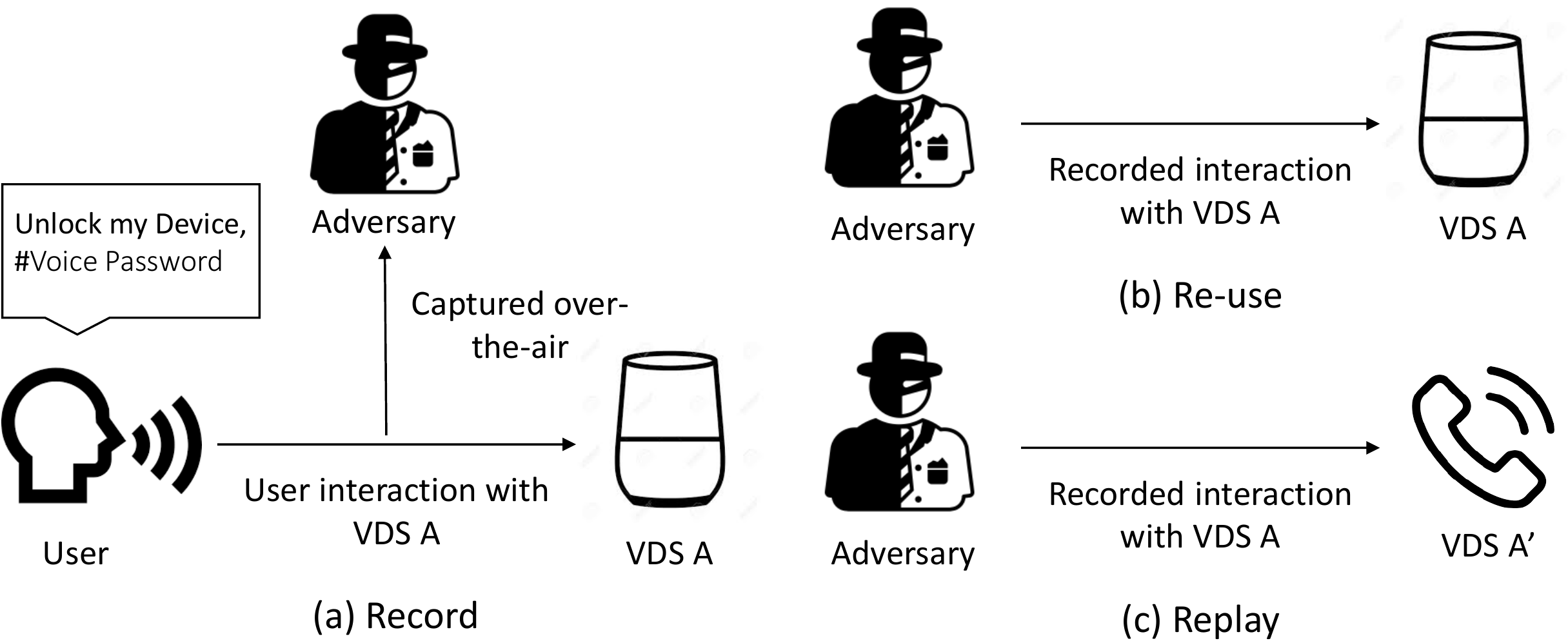}
  \caption{Difference between re-use and replay attacks. In both cases, an adversary records a user’s interaction with a voice driven service (VDS) ``A'', over-the-air as shown in (a). In case of re-use (b), the recorded interaction is used to attack the same type of VDS A, while in case of replay (c), the interaction can be used to attack the same or different type of VDS A’.}
  \label{fig:firstarch}
\end{figure}

Voice passwords, however, do not provide security against voice spoofing techniques such as speech replays~\cite{gong2018protecting,pradhan2019combating,leng2020replay} and synthesis~\cite{synthesisattack}. While there are known practical limitations in conducting speech synthesis attacks (e.g., the attack proposed in~\cite{synthesisattack} requires 24 hours of high quality training data), it is easier to conduct replay attacks in practice. Broadly, there are two categories of defense mechanisms that aim to prevent speech replays. The first class of methods use machine learning to determine whether a speech sample is produced by a human or replayed using a playback device~\cite{chen2017you,wang2020differences}. The second category of techniques use external devices (e.g., a vibration sensor) to check if the speech is produced by a human in real-time~\cite{feng2017continuous}. For machine learning techniques, real world performance usually depends on training data and its relevance to the test environment. A state-of-the-art method Void~\cite{ahmed2020void}, as an example, has significantly different equal error rates (EER), 0.3\% and 11.6\%, on two speech datasets collected in two different environments. Another practical limitation is unreliability of playback device signatures used by machine learning techniques~\cite{wang2020differences}. In addition, for methods that use extra hardware, there are usability and cost implications for practical deployment.

\begin{figure}[t]
\centering
\includegraphics[width=0.8\linewidth]{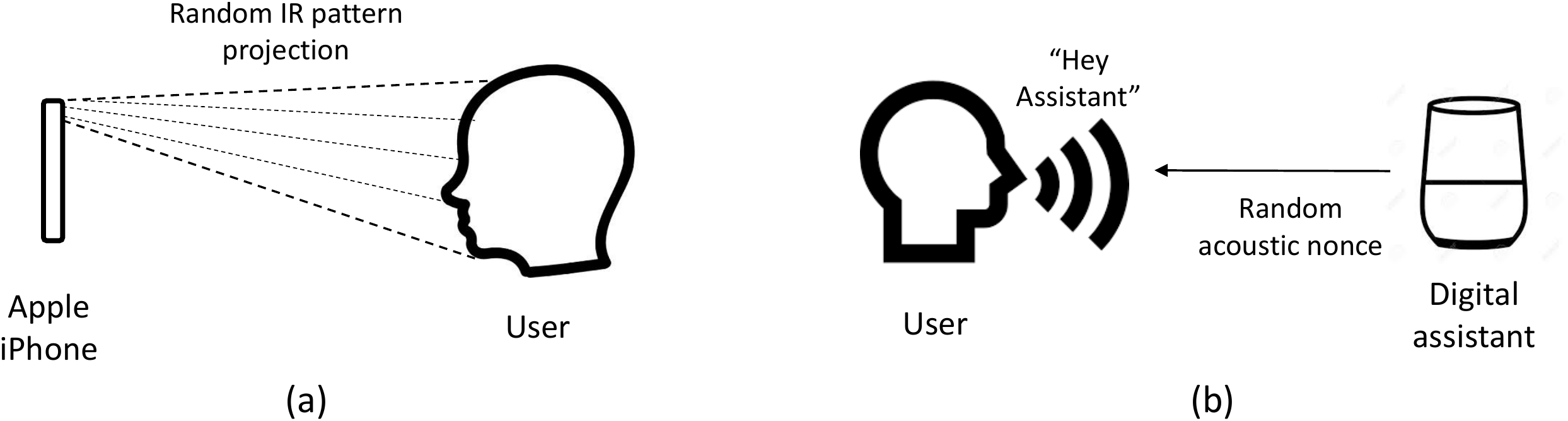}
  \caption{(a) Face ID's device-specific random IR pattern projection at the time of user authentication for preventing re-use~\cite{faceID_security}. Analogous to (a), the proposed security overlay (b) uses a random acoustic nonce at the time of user interaction to prevent speech re-use in voice-driven systems.}
  \label{fig:facial}
\end{figure}

In this work, we aim to address a subset of replay attacks, called \textit{speech re-use} attacks. We define speech re-use attacks to be specific to a certain type of VDS. As shown in Figure~\ref{fig:firstarch}, let A and A' be the VDS types for the capture and attack steps, respectively. Replay attacks typically consider a variety of VDS types for capture and attack steps (A = A' and A $\neq$ A'), while re-use attacks are a subset of replay attacks where the VDS types in consideration for the capture and attack steps is similar (A = A'). The proposed security overlay prevents re-use attacks on protected VDS. For example, if the security overlay is integrated with each Alexa's VDS, it can prevent speech captured from user interaction with one Alexa's VDS from being re-used on another Alexa's VDS. This is similar to the device specific random pattern used in Face ID to counter digital and physical spoofs~\cite{faceID_security} (see Figure \ref{fig:facial}).

We present \tool\footnote{Named after Aelous, the "Keeper of the Winds" in Greek mythology.}, a security overlay that proactively embeds a dynamic acoustic nonce in the voice channel via the microphone at the time of user interaction, and detects the presence of the embedded nonce in the speech recorded by the speaker to ensure speech freshness. \tool\ is designed as a software solution, and can be integrated by a vendor with any closed-loop \vcs that involves real-time user interaction. It is useful for providing additional security in critical applications such as user authentication and payments. \tool\ does not require any extra hardware and works with in-built speaker and microphone in consumer products. \tool\ is secure by design against attacks that remove the embedded nonce for speech re-use. It uses Frequency-hopping spread spectrum (FHSS)~\cite{cvejic2004algorithms}~\cite{cvejic2004spread} technique for dynamic acoustic nonce embedding. Similar techniques have been previously used to watermark audio recordings for copyright purposes~\cite{swanson1998robust}~\cite{bassia2001robust}. However, their application in offline audio watermarking is fundamentally different from \vcs operational setting where the length and content of the input speech is unknown apriori. \tool\ is designed for real-time over-the-air use without prior knowledge of the input speech.

\tool\ addresses two key practical challenges. The first challenge is to reliably embed an acoustic nonce over-the-air and retrieve it successfully from the recorded speech. Over-the-air acoustic nonce propagation is affected by factors such as background noise and distance between microphone and loudspeaker~\cite{chenmetamorph}. The latter, for instance, results in distorted and attenuated speech signal which increases the likelihood of errors while extracting the embedded acoustic nonce. The second challenge is to embed the acoustic nonce such that it is imperceptible to a \vcs user. Achieving imperceptibility is non-trivial in light of the first challenge. This is because an important consideration for reliability is embedding acoustic nonce of certain strength which, in turn, makes the acoustic nonce perceptible. Therefore, we model acoustic nonce generation as an optimization problem and compute the set of optimal parameters, e.g., nonce's operating frequency, amplitude, bitrate in different environments using differential evolution.

We evaluate \tool\ in three different environments and show that it works reliably upto a range of 4 m.
Additionally, we conduct a user study involving 120 subjects (approved by the institutional human study review board) to evaluate the imperceptibility of the embedded nonce. The results of the study show that majority of users find the embedded acoustic nonce to be either imperceptible or non-disruptive.

\noindent In summary, the contributions of this work are as follows:

\begin{itemize}
    \item Design of a \vcs security overlay called \tool\ to ensure freshness of input speech without any extra hardware dependency. 
       
     \item Modeling acoustic nonce propagation as an optimization problem to address i) reliability, i.e. ensuring successful embedding and retrieval of dynamic acoustic nonce without impacting VDS functionality, and (ii) imperceptibility, i.e. to have minimal impact on VDS users' experience. 
    
    \item Comprehensive real-world evaluation to show that \tool\ can  work effectively (0.5\% false reject rate (FRR) and 0\% false accept rate (FAR) in detecting speech re-use) up to a range of 4 meters in three different environments.

    \item User study with 120 subjects to demonstrate that the embedded acoustic nonce does not degrade user experience.
    
\end{itemize}

\section{Background and Related Work}

\subsection{Example Use Cases}
\label{sec:motivate}
\smallskip\noindent\textbf{Voice-driven Payments.}
Voice-driven services are being increasingly used for payments, especially in retail environments~\cite{vpsurvey}. Amazon and Exxon recently announced voice-driven payments at over 11,500 gas stations in the US~\cite{gasstationpay}. A recent user study~\cite{paysafe} indicates that while most users are comfortable conducting low-value purchases such as ordering a meal and shopping for groceries using voice-driven payment services, the majority of them do not have sufficient confidence in conducting high value purchases due to security concerns.

Most voice-driven payment services authenticate users through speaker verification. However, they often lack adequate protection mechanisms against speech re-use attacks where an attacker records a user interaction and re-uses it to attack the service either on-site or remotely. This is analogous to card skimming~\cite{scaife2018fear} where an attacker re-uses a victim's credit card information. \tool\ can be deployed to prevent speech re-use attacks in voice-driven payments.

\smallskip\noindent\textbf{Workplace Automation.}
A variety of digital assistants are being used for unsupervised or semi-supervised interactions in workplace environments~\cite{alexabusiness}. In addition to accurate speaker recognition (identification and/or verification), it is important to have adequate protection mechanisms against speech re-use to protect sensitive resources in such environments~\cite{workplaceva}. \tool\ can be incorporated in workplace digital assistants for this purpose.

\subsection{Related Work}

\smallskip\noindent\textbf{Speech Replay Detection}
Existing research related to speech replay detection can be classified into two categories, software-based and hardware-based methods. Software-based methods use machine learning to determine whether the input speech is produced by a human or replayed using a recording device~\cite{chen2017you}~\cite{ahmed2020void}~\cite{kinnunen2017asvspoof}. Chen et al.~\cite{chen2017you} present a method that obtains 0\% EER for distance (between user and \vcs input device) of only a few centimeters which limits its applicability in practice. Ahmed et al.~\cite{ahmed2020void}'s method, on the other hand, is reported to work well upto a distance of 2.6 meters. However, the EER varies significantly in different environments with it being as low as 11.6\%.
One of the best performing methods ~\cite{kinnunen2017asvspoof} reports EER of 6.7\% on ASVSpoof 17 database. A caveat, however, is that these methods aim to address general replay attacks. In contrast, \tool\ aims to prevent speech re-use and obtains 0\% EER upto a distance of 4 meters in three different environments.

The second category of approaches use additional hardware (e.g., a vibration sensor on user's neck) to check that speech is produced in real-time by a user~\cite{feng2017continuous}~\cite{pradhan2019combating}. Such approaches, in general, outperform machine learning-based methods. For example, Feng et al.~\cite{feng2017continuous} report 97\% successful detection rate, on average, with no strict assumption on distance. However, there are two major practical limitations using the aforementioned approaches for detecting speech re-use, (i) requirement of extra hardware which has cost implications, and (ii) inconvenience to users. \tool\ only uses the built-in speaker and microphone in smart devices yet achieves high speech re-use detection rate.

\smallskip\noindent\textbf{Audio Watermarking.}
Audio watermarking is widely used for multimedia copyright protection~\cite{swanson1998robust,seok2001audio}. Both time domain~\cite{bassia2001robust} and frequency domain~\cite{cox1996secure,neubauer1998digital} watermarking techniques have been proposed in the literature. Typically, audio watermark embedding and detection is self-synchronized~\cite{wu2005efficiently}. Advanced watermarking methods use spread-spectrum modulation~\cite{swanson1998robust} to prevent watermark removal and provide imperceptibility~\cite{cox1996secure,kirovski2003spread}. While useful, these techniques cannot be directly used in \vcs operational setting. They are designed for fixed length offline audio files, e.g., music recordings in CDs~\cite{swanson1998robust}, and do not consider the impact of environmental factors, e.g., bitrate variability and background noise, in over-the-air transmission. Such environmental factors result in a dynamic and lossy acoustic environment in turn making watermark embedding and retrieval significantly challenging. Further, they are not designed for real-time speech and speaker recognition systems where input speech is unknown apriori and can be of variable length. \tool\ is designed to take these pragmatic considerations into account. It is able to successfully embed and retrieve acoustic nonce in the voice channel without degrading user experience in practice.

\section{Proposed Security Overlay}
\label{sec:design}

\begin{figure}[t]
\centering
 \begin{subfigure}[b]{0.49\textwidth}
         \centering
            \includegraphics[width=1\linewidth]{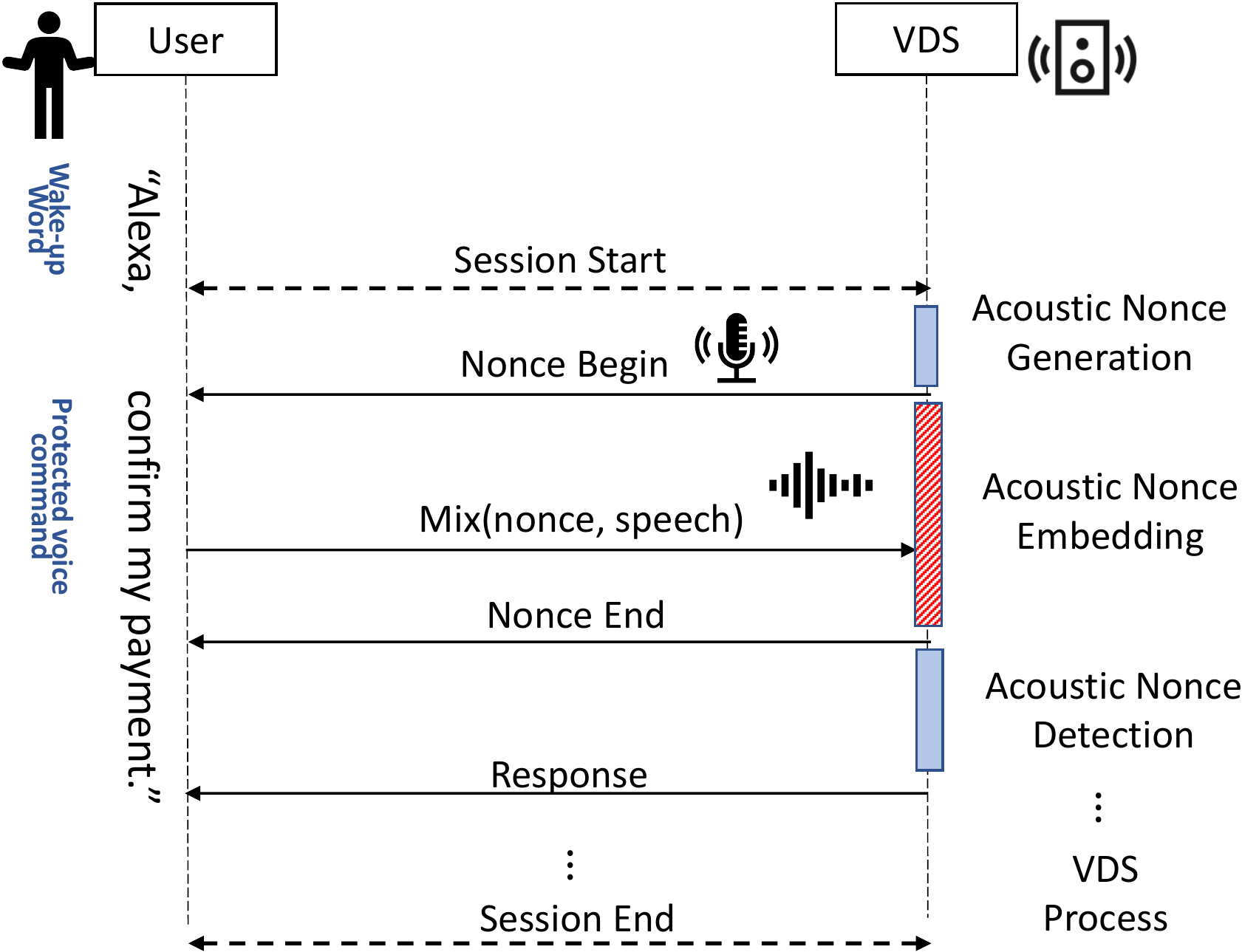} 
            \caption{}
         \label{fig:overview3_a}
\end{subfigure}
\hfill
 \begin{subfigure}[b]{0.49\textwidth}
         \centering
            \includegraphics[width=1\linewidth]{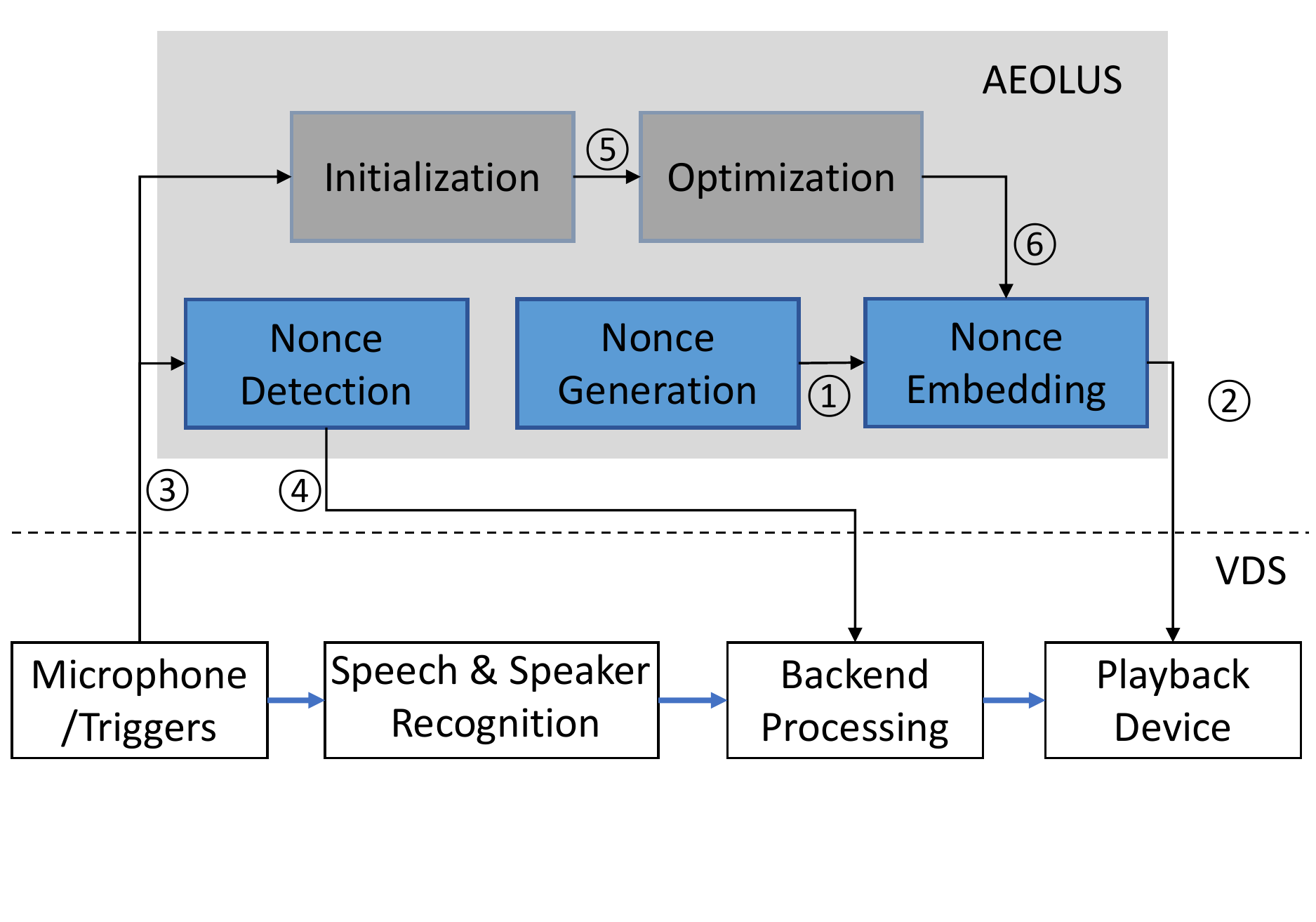} 
            \caption{}
         \label{fig:overview3_b}
\end{subfigure}
  \caption{Use of the proposed security overlay to prevent speech re-use in voice-driven services. (a) At the time of user interaction, an acoustic nonce is generated and embedded in over-the-air channel, and its presence is detected in the recorded speech to ensure speech freshness. (b) \tool{-enabled} \vcs components with the associated data flows. Shown in blue and grey are the components used to prevent speech re-use.}
  \label{fig:overview3}
\end{figure}

\subsection{Threat Model}
The following assumptions are made while designing \tool\ to address the speech re-use attack scenario. The hardware device (loudspeaker and microphone) that is used by a user to interact with \vcs is trusted and secure, and an adversary cannot disable the speaker or the microphone. The proposed overlay can be used by a \vcs to generate, embed, and detect acoustic nonce. At the time of user interaction, the method used to trigger a \vcs,  e.g., words or phrases like ``Alexa'' and ``Hey, Siri'' are not protected but can initiate the proposed overlay to prevent re-use of voice commands and confidential passphrases. For example, to prevent re-use of a user's voice password to access recent account transactions when the user is interacting with a voice banking \vcs~\cite{paypalpin}.

An adversary can use a commodity recording device to record prior user interactions with ``any'' \tool-enabled \vcs, and re-use the recorded interaction to attack any \tool-enabled \vcs. Attack scenarios where an adversary can record or access clean user speech samples (i.e., recordings without embedded acoustic nonce) to launch replay or speech synthesis attacks are considered out of scope and are not addressed. The key underlying assumptions are that (i) users typically interact with \vcs using specific commands (e.g., a user-defined passphrase~\cite{bixbypassword}) and these commands can be protected with the proposed overlay, and (ii) it is difficult to record clean samples of these commands from interaction with unprotected devices or in human conversations. However, this does not limit the use of \tool\ in conjunction with other defense mechanisms for replay or speech synthesis attacks.

\subsection{Core Components}
\label{subsec:goals}

To prevent speech re-use in \vcs, \tool\ uses three core components, nonce generation, nonce embedding and nonce detection. 


\smallskip\noindent\textbf{Nonce Generation.}
The nonce generation module is invoked upon initiation of a new user interaction session (see Figure~\ref{fig:overview3_b} \textcircled{1}). Akin to nonce-based authentication~\cite{tsai2009efficient,bellare1995provably}, the module generates a digital nonce (a random number of a length $L$) that is valid for the current session. Nonce length $L$ should be sufficiently large to prevent use of duplicate nonces in different sessions. Given limited bandwidth of over-the-air channel and estimated user base for popular \vcs (e.g., Alexa skill has 300,000 daily users~\cite{alexadailyuser}), the minimum permitted value of $L$ is set to 32 bits. 32-bit long nonce is sufficient to protect short speech samples (e.g., a voice passphrase) in a particular session. Nonce repetition is permitted to protect longer speech samples during the session. The nonce $\delta$ for session $S$ is stored in a set $S_{set}$. $S_{set}$ is subsequently used by the nonce detection module to determine if the embedded nonce is current.

\smallskip\noindent\textbf{Nonce Embedding.}
 Post nonce generation, an encoding algorithm, Binary Frequency Shift Keying (BFSK)~\cite{park2010underwater,truax2001acoustic}, is used to encode the generated digital nonce as an acoustic nonce. Each bit is independently encoded as a cosine wave $\omega_i$. Bit ``0'' is encoded as a 1kHz cosine wave, and bit ``1'' as a 5 kHz cosine wave. All cosine waves in the set $\omega$ are concatenated to yield the acoustic nonce $\delta$. The acoustic nonce $\delta$ is embedded in over-the-air channel using \vcs playback device ( Figure~\ref{fig:overview3_b} \textcircled{2}). An important parameter for nonce embedding is the time duration of the embedded acoustic nonce relative to the time duration of user interaction. Using a real-world dataset with 27,000 samples~\cite{gong2019remasc}, we estimate that the average user interaction is 3 sec. As an example, let the minimum duration of embedded nonce be half of the average user interaction. In this case, the acoustic nonce embedding would require a bitrate of 14 bits/sec, and the maximum duration of each component cosine wave $w_i$ can be 42 ms. Note that this estimate does not consider the idle time in user interaction. Additionally, to prevent an adversary from obtaining clean speech sample by removing the embedded nonce, the nonce embedding module leverages Frequency-hopping Spread Spectrum (FHSS)~\cite{cvejic2004spread,cvejic2004algorithms}. FHSS is a well-known technique that provides high robustness against standard embedded signal removal attacks. For high robustness, the module uses FHSS with a set of different frequencies and periodically selects the operating frequency at random.

\smallskip\noindent\textbf{Nonce Detection.}
Once user interaction ends, the recorded audio is processed using the nonce detection module (Figure~\ref{fig:overview3_b} \textcircled{3}). The module decodes the acoustic nonce using BFSK, and checks that (i) the nonce is completely decoded, and (ii) the decoded nonce is current. The recorded speech is subsequently processed using standard \vcs modules (Figure~\ref{fig:overview3_b} \textcircled{4}). If nonce detection fails or if the decoded nonce has a high Bit Error Rate (BER), a recourse action (e.g. a retry or reject) can be taken. Furthermore, if speech is reused by an adversary and both the current nonce and the obsolete nonce from re-used speech are detected, information about re-used speech such as when and where it was recorded can be determined from $S_{set}$. Another scenario can be signal interference between the current and obsolete nonce resulting in inappropriate decoding result. In such case, pertinent recourse action can be taken.
\section{Practical Realization}
\label{sec:new design}

\subsection{Challenges}
In this section, we discuss the two key challenges while implementing \tool\ in practice: reliability and imperceptibility.

\smallskip\noindent\textbf{Reliability.}
The first challenge is to reliably embed an acoustic nonce in over-the-air channel, as well as to accurately detect nonce from the recorded speech. Because over-the-air acoustic nonce transmission is impacted by factors such as background noise and distance between microphone and playback device, it is difficult to achieve this in operational settings. It is also important to determine if any previously embedded acoustic nonce is present in the recorded speech to ascertain speech re-use. The metric used to measure reliability is the bit error rate (BER) between the embedded and detected nonce. Recall that BFSK is used for acoustic nonce embedding and detection. Under the assumption that only additive white Gaussian noise (AWGN) is present in over-the-air channel, BER can be computed using the complimentary error function $erfc()$~\cite{kumar2010comparison} as follows:

\begin{equation}
\label{eqn:ber}
 BER = \frac{1}{2} erfc(\sqrt{E_b / N_0}) 
\end{equation}

\noindent Note that BER computation in Eqn.~\ref{eqn:ber} involves the normalized per bit signal-to-noise ratio ${E_b/N_0}$~\cite{johnson2006signal} which depends on the frequency $f$ and amplitude $\alpha$ of the carrier wave. However, other types of noises besides AWGN are typically present in over-the-air channel. Hence, we conduct Room Impulse Response (RIR) simulation experiments to study the impact of carrier wave amplitude and frequency on BER.

\smallskip\noindent\textbf{Imperceptibility.} The second challenge is to ensure that the embedded acoustic nonce does not degrade \vcs user experience. For this, the acoustic nonce should be as imperceptible as possible to a \vcs user. The nonce embedding and generation modules presented earlier do not adequately address this challenge because (i) they do not optimize any objective metric for imperceptibility, and (ii) they do not account for dynamic and noisy environments where it is difficult to achieve both reliability and imperceptibility simultaneously. For measuring imperceptibility, Sound Pressure Level (SPL) is computed using the following equation:

\begin{equation}
\label{eqn:soundpressure}
	SPL = 2 {\pi}^2{f}^2{\alpha}^2\rho{c}/{v}
\end{equation}

\noindent Here, $f$ represents the frequency and $\alpha$ is the amplitude of the carrier wave. $\rho$ represents the density of the transmission channel (e.g., over-the-air channel), and $c$ denotes the speed of acoustic transmission. Given that the average SPL for human speech is 60dBm~\cite{human_normal_hearing}, $f$ and $\alpha$ should ideally ensure that SPL is below this threshold for imperceptibility. Like Equation~\ref{eqn:ber}, while this equation provides a theoretical basis to understand SPL, we study how the aforementioned parameters impact SPL using RIR simulation.

\subsection{Key Parameters}
There are two key parameters that impact reliability and imperceptibility: (i) acoustic nonce generation and embedding parameters that include frequency and amplitude of the carrier wave used for acoustic nonce generation and bitrate used for acoustic nonce encoding, and (ii) environmental parameters that include the distance between the microphone and user, the room size, and background noise among others. To understand the impact of the these parameters on reliability and imperceptibility, we implement \tool\ as a prototype and setup an RIR environment called MCRoomSim in MATLAB~\cite{wabnitz2010room}. RIR is a commonly used tool for estimating the performance of signal processing applications in over-the-air channel~\cite{chenmetamorph,abdullah2019practical}.

\begin{figure*}[t]
\centering

 \begin{subfigure}[b]{0.3\textwidth}
         \centering
            \includegraphics[width=1\linewidth]{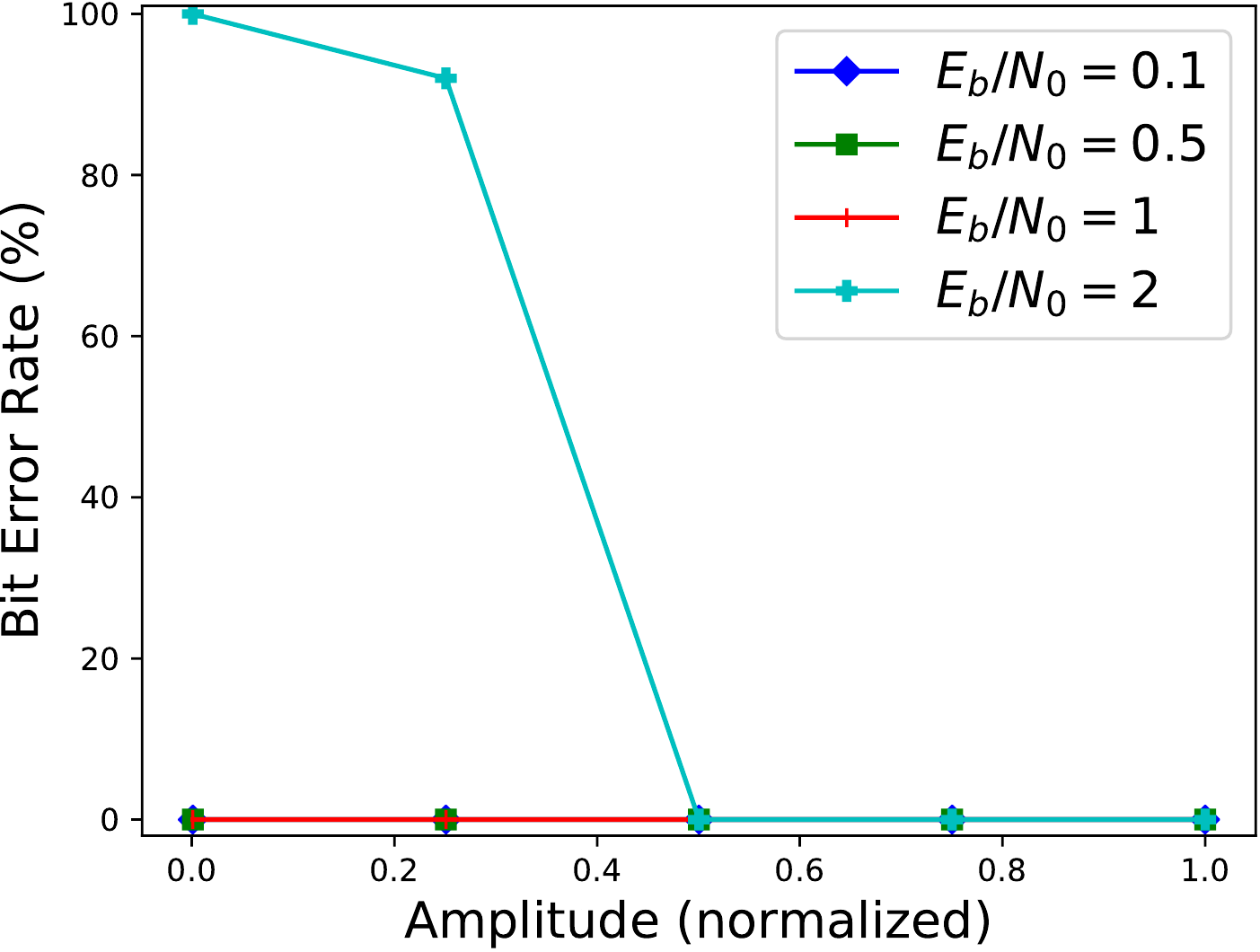}   
        \caption{Frequency-range=4-5kHz, hopping = 2, bitrate = 30 bits/sec}
         \label{fig:BER_a}
\end{subfigure}
\hfill
 \begin{subfigure}[b]{0.3\textwidth}
         \centering
            \includegraphics[width=1\linewidth]{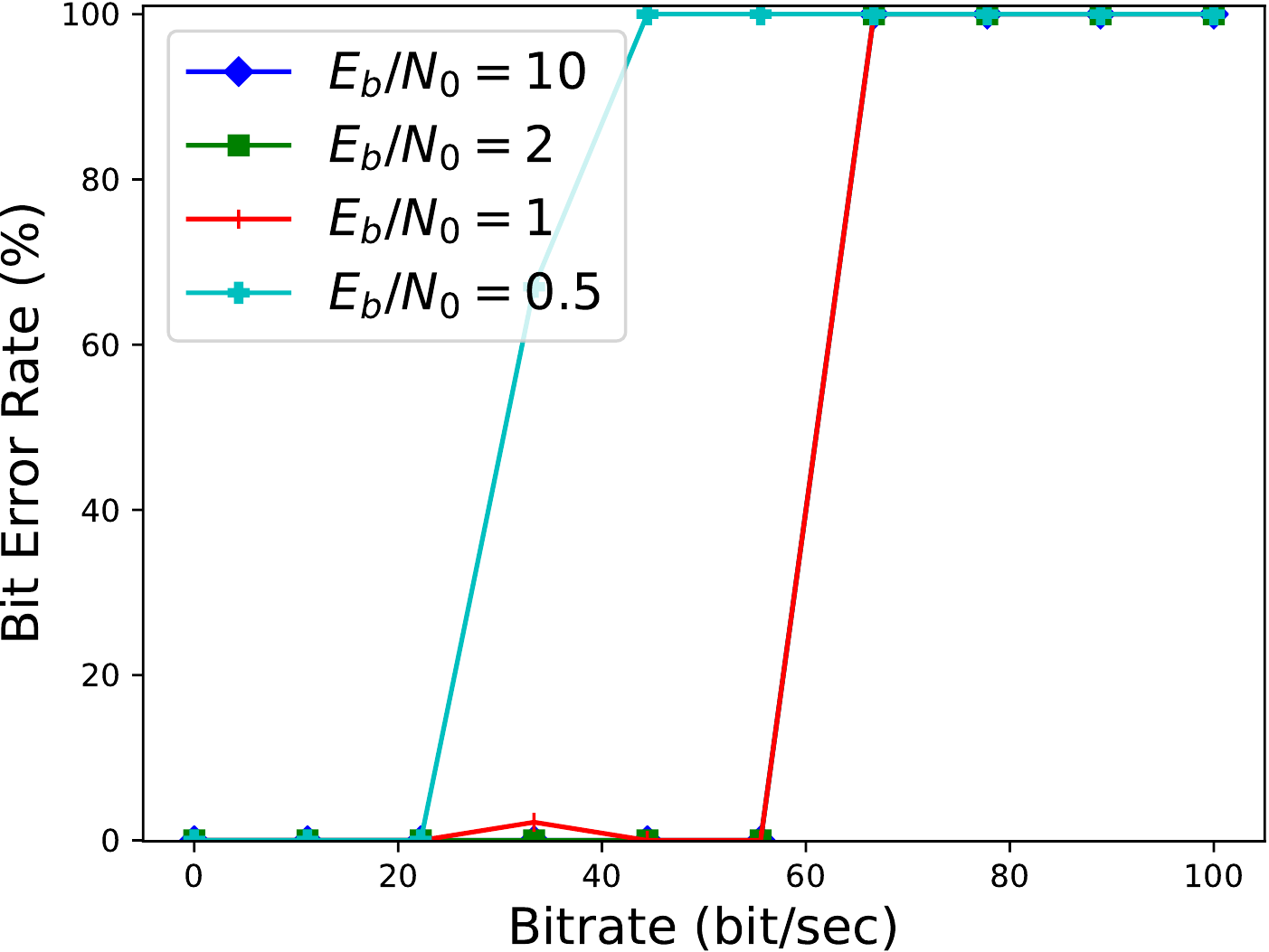}   
        \caption{Frequency-range=4-5kHz, hopping=2, amplitude = 1}
         \label{fig:BER_b}
\end{subfigure}
\hfill
 \begin{subfigure}[b]{0.3\textwidth}
         \centering
            \includegraphics[width=1\linewidth]{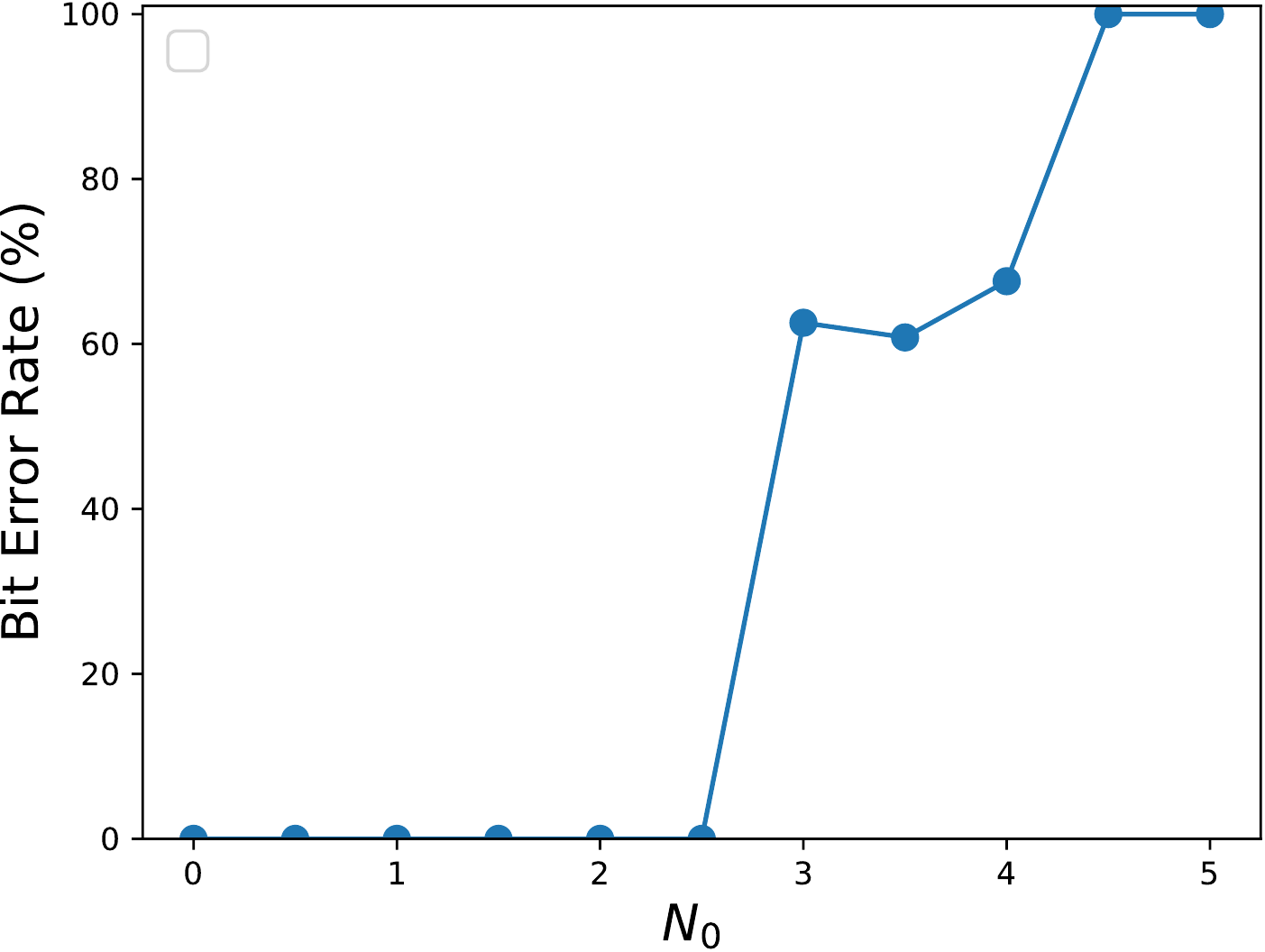}   
        \caption{Freq-range=4-5kHz, hopping=2, bitrate=30 bits/sec, amplitude=1}
         \label{fig:BER_c}
\end{subfigure}
  \caption{Variation of bit error rate (\%) with: (a) amplitude of carrier wave, (b) bitrate used for encoding acoustic nonce, (c) background noise levels when embedding the acoustic nonce. $E_b/N_0$ is calculated in decibels. The distance between user (simulated using a speaker) and the microphone is set to 1 m.}
  \label{fig:sim_freq}
  \vspace{-4mm}
\end{figure*}

\subsubsection{Nonce Generation and Embedding Parameters}\hfill

\smallskip \noindent \textit{Amplitude.} According to RIR simulation, a lower amplitude yields lower $E_b/N_0$ ratio and consequently, higher BER (see Figure \ref{fig:BER_a}). This is consistent with Equation~\ref{eqn:ber}. Furthermore, as shown in Figure~\ref{fig:sim_spl}, SPL is proportional to the amplitude of the carrier wave. 

\smallskip \noindent \textit{Frequency.}
BER and SPL are evaluated in the frequency range at which typical speaker recognition systems operate (5 Hz - 8 kHz; sampling rate = 500 Hz). The other parameters are fixed as follows: ${E_b/N_0}$ = 1, frequency hopping = 2, amplitude (normalized) = 1, bitrate = 30 bits/sec, distance = 1 meter. It is observed that the carrier wave frequency neither impacts BER nor SPL. The later is in contrast to existing research~\cite{suzuki2004equal,wiki_elc} which shows that the frequency of an acoustic wave affects the loudness as perceived by humans in a non-linear manner. Hence, the frequency to SPL mapping from IoSR toolkit~\cite{githubmapping} is used as reference.

\smallskip \noindent \textit{Bitrate.}  Figure~\ref{fig:BER_b} shows the impact of acoustic nonce bitrate on BER. BER increases significantly as bitrate increases beyond 20 bits/sec. One of the causes is asynchronous signal overlapping in the time domain when two bits are transmitted too close to each other.

\subsubsection{Environmental Parameters}\hfill

\begin{figure}[t]
\centering
 \begin{subfigure}[b]{0.40\textwidth}
         \centering
            \includegraphics[width=1\linewidth]{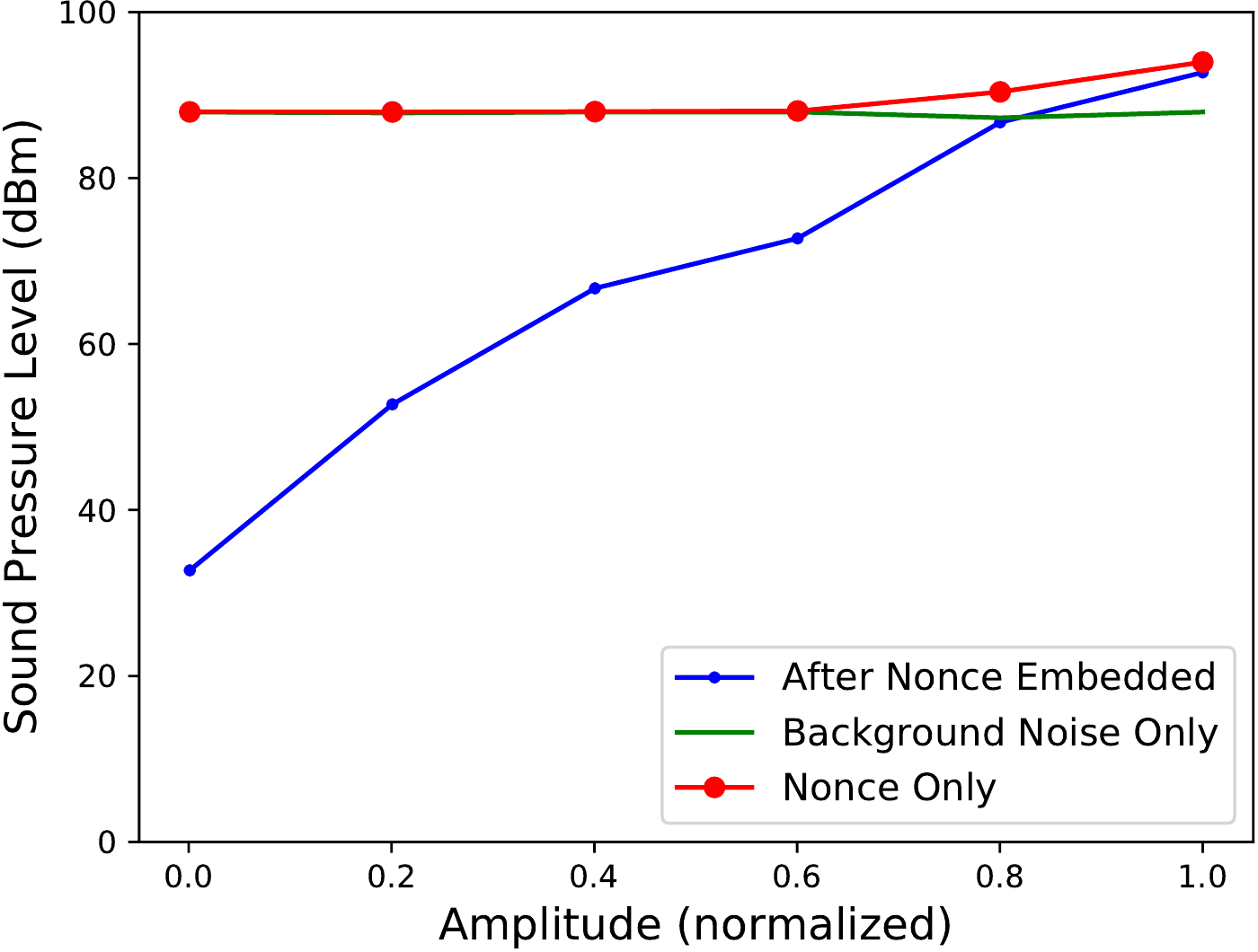}   
        \caption{$E_b/N_0$ = 1, carrier frequency range = 4-5kHz, hopping=2, bitrate = 30 bits/sec, distance = 1m}
         \label{fig:SPL_a}
\end{subfigure}
\hfill
 \begin{subfigure}[b]{0.45\textwidth}
         \centering
            \includegraphics[width=1\linewidth]{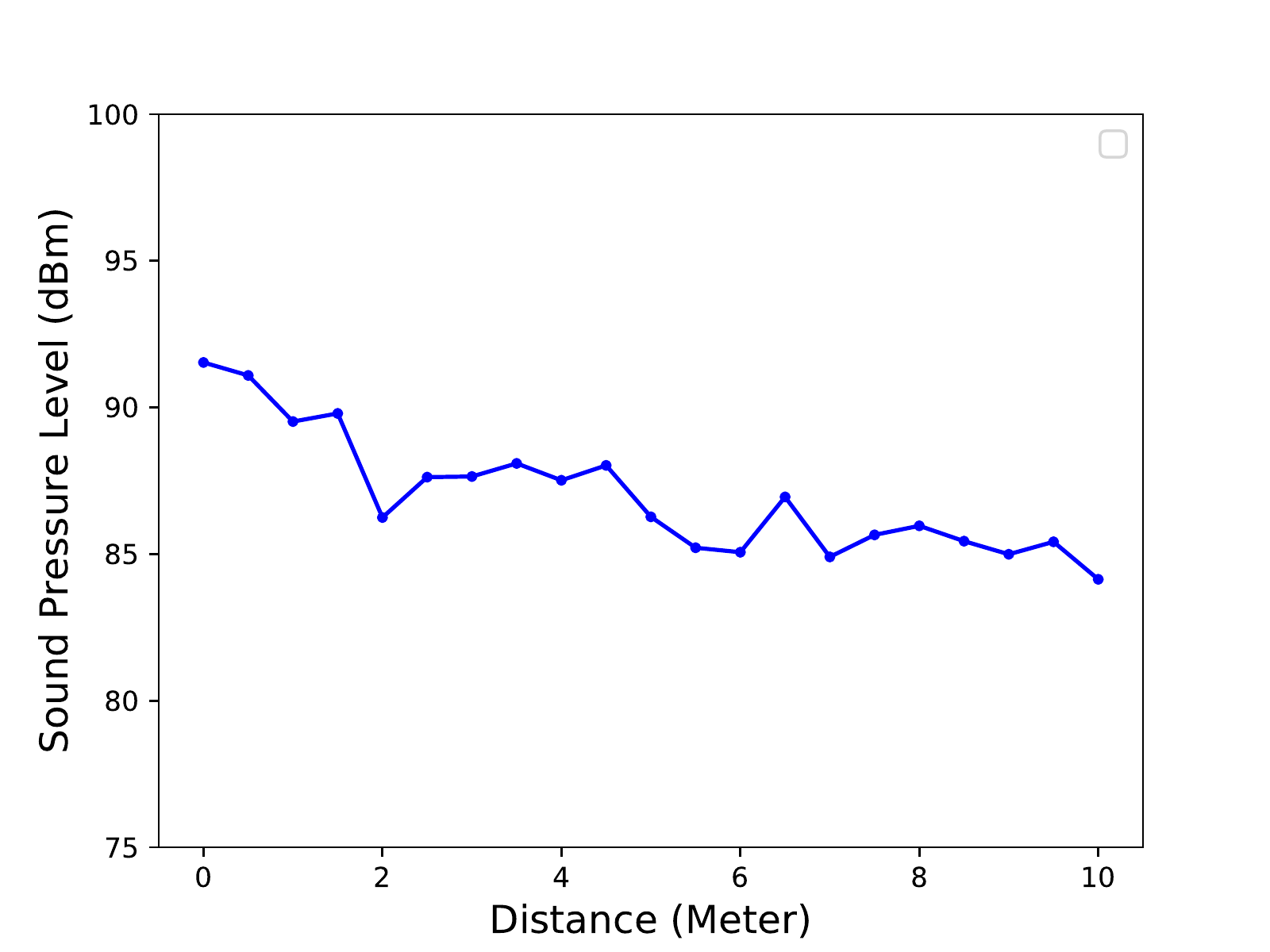}   
        \caption{$E_b/N_0$ = 1, carrier frequency range = 4-5kHz, hopping=2, bitrate = 30 bits/sec, Amplitude (normalized) = 1}
         \label{fig:SPL_b}
\end{subfigure}

\caption{Variation of sound pressure level (dBm) with (a) amplitude of carrier wave, and (b) distance between user (simulated using a speaker) and the microphone.}
 \label{fig:sim_spl}
\end{figure}

\smallskip \noindent \textit{Frequency Shifting.}
Over-the-air acoustic nonce transmission induces differences in the received frequencies and the transmitted frequencies called \textit{frequency shifting}. Frequency shifting can also occur because of hardware limitations, e.g., a microphone with a 1.8 kHz frequency limit being used to capture a 2 kHz wave. Since the received frequency is used to decode the acoustic nonce, frequency shifting needs to be accounted for by \tool. We investigate frequency shifting in the simulation setting (operating frequency range: 500 Hz-8 kHz, cardioid loudspeaker type, dipole microphone type) and determine the frequency shifting range to be +/- 15 Hz.

\smallskip \noindent\textit{Distance.}
Distance between a \vcs user and \vcs input impacts imperceptibility of the acoustic nonce. Figure~\ref{fig:SPL_b} indicates that this distance is inversely proportional to SPL.

\smallskip \noindent \textit{Background Noise.}
Figure~\ref{fig:BER_c} and  Equation~\ref{eqn:ber} suggest that the background noise is proportional to BER. In an environment with high background noise, the nonce generation and embedding parameters need to be configured appropriately (e.g., decrease bitrate, increase amplitude) to limit the BER.

\subsection{Mathematical Formulation}
\label{sec:optfor}
To address reliability and imperceptibility simultaneously, we model over-the-air acoustic nonce transmission as an optimization problem. Let a user's recited speech be denoted by $x$, and the acoustic nonce embedded by \tool\ be $\delta$. Also, let the function that accounts for variations induced by over-the-air channel, e.g., due to acoustic signal air absorption and distance, be $h(\cdot)$. The speech recorded by the microphone is thus represented as $h(x)$. When $\delta$ is embedded at the time of user interaction with the \vcs, $x$ mixes with $\delta$. As a result, the microphone records $h(\textbf{x}* \delta)$, where $*$ denotes signal convolution between $x$ and $\delta$.

Assume that \tool\ detects nonce $\delta{'}$ in $h(x * \delta)$. Let us define an objective function $f(\cdot)$ denoting BER that takes the parameter set $\mathcal{P} = \{<f,\alpha,\tau>, ...\}$ as input, and a function $g(\cdot)$ denoting SPL. The goal is to find the optimal parameter set $\mathcal{P}$ that minimizes $f$ (for reliability) subject to the constraint that the SPL difference between speech with and without embedded nonce is less than or equal to a threshold $\theta$ (for imperceptibility):

\begin{equation}
\label{eqn:optim}
 \arg \min_{\mathcal{P}} f(\mathcal{P}) \\
 \text{ subject to } g(h(\textbf{x}* \delta)) - g(h(\textbf{x})) \leq \theta
\end{equation}

\begin{algorithm}[t]
\SetKwInOut{Input}{input}
\SetKwInOut{Output}{output}
\small
\SetAlgoLined
\textbf{initialization:} \\
1. Create candidate population:$\mathcal{P} = \{P<f,\alpha,\tau>, ...\}$\\
2. Initialize: $\mathcal{P}_{output} =\{\}$, $i = 0$, $MAX_i = 100$, $MAX_{output} = 20$, $\beta = 0.5$\\

2. Acquire environmental parameters: $d$, $N_0$

3. Set up Goals:$BER < 10^{-9}$, $SPL^{'} - SPL \leq \theta$\\

\textbf{optimization:} \\
\While{$i<= MAX_i$, $\mathcal{P}_{output}.Size <= MAX_{output}$ }{ 
\ForEach {individual $P_i (i = 1, ...) \in \mathcal{P}$} {
\If{$P_i$ satisfies the Goals}
{$\mathcal{P}_{output} \leftarrow P_i$ }
Create candidate $C$ from parent $P_i$, $d$, $N_0$.\\
\eIf{the candidate is better than the parent}
{$P_{feasible} \leftarrow$ Candidate $C$ \\
Candidate $C$ replaces the parent}
{the candidate is discarded}
} Randomly select next $P$ in $\mathcal{P}$.}

\textbf{output:} $\mathcal{P}_{output}$
 \vspace{2mm}
\hrule \vspace{0.8mm}
\textbf{Candidate Creation:} 
\vspace{0.8mm}
\hrule
\vspace{2mm}
\Input{Parent $P_{par}$, $d$, $N_0$, $\beta$}
Randomly select individuals $P_{1}$ , $P_{2}$  from $\mathcal{P}$\\
Calculate candidate $C$ as $C = P_{par} + \beta (P_{1} - P_{2} )$,\\
where $\beta$ is a scaling parameter decided by $d$ and $N_0$. \\
\Output{Candidate $C$}
 \caption{Computing Optimal Parameters}
 \label{alg:de}
\end{algorithm}

\subsection{Computing Optimal Parameters}
\label{subsec:optimization}
 
\smallskip \noindent \textbf{Initialization.}
The environmental parameters are first initialized (Figure~\ref{fig:overview3_b} \textcircled{5}) using the \vcs input. The background noise energy is calculated by inspecting the ambient sound. Frequency shifting is calculated by transmitting and decoding a nonce that covers the working frequency range (500 - 8 kHz, sampled every 500 Hz).  It is assumed that the distance between the user and \vcs input device is set by the \vcs device vendor.
 
\smallskip \noindent \textbf{Optimization.}
Computing the gradients of the optimization parameters directly using gradient-based methods is inefficient~\cite{chen2019real}. This is because the \vcs{'s} acoustic environment is dynamic (e.g., due to occlusions and varying background noise levels). In addition, the formulated optimization is a multi-dimensional problem with considerations of both reliability and imperceptibility. Hence, we use differential evolution (DE)~\cite{das2010differential,storn1997differential} which is a gradient-free method. DE generates diverse candidate solutions in each iteration which avoids local minima and aids in convergence to a global minima. It does not require apriori knowledge of the underlying acoustic encoding method or acoustic environment. Algorithm~\ref{alg:de} summarizes the use of DE for obtaining optimal parameters.
The initial set of candidate parameters $\mathcal{P}$ includes feasible frequency $f$, amplitude $\alpha$, and bitrate $\tau$, estimated using RIR simulation. Once \tool\ is deployed, the initialization module first estimates the environmental parameters, e.g., $N_0$ and $d$. Subsequently, the optimization module uses the pre-calculated RIR parameters for efficient convergence to optimal parameters (see Section~\ref{subsec:eval_performance}).

\section{Experimental Evaluation}
\label{sec:eval}

\begin{figure}[t]
        \centering
        \begin{subfigure}[b]{0.244\textwidth}
            \centering
            \includegraphics[width=\textwidth]{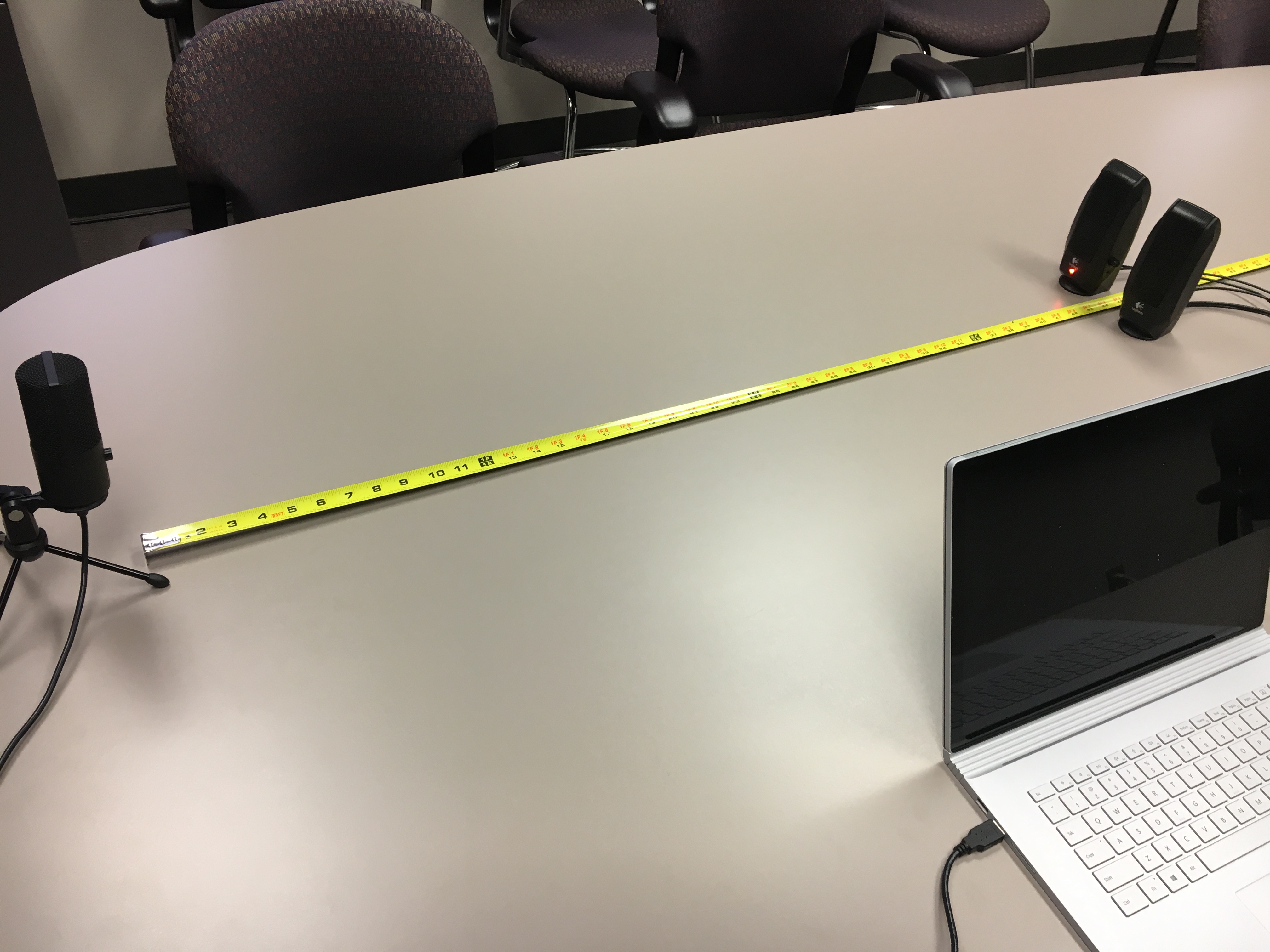}
            \caption[Conference Room.]%
            {{\scriptsize Conference Room.}}    
            \label{fig:mean and std of net14}
        \end{subfigure}
        \hfill
        \begin{subfigure}[b]{0.244\textwidth}  
            \centering 
            \includegraphics[width=\textwidth]{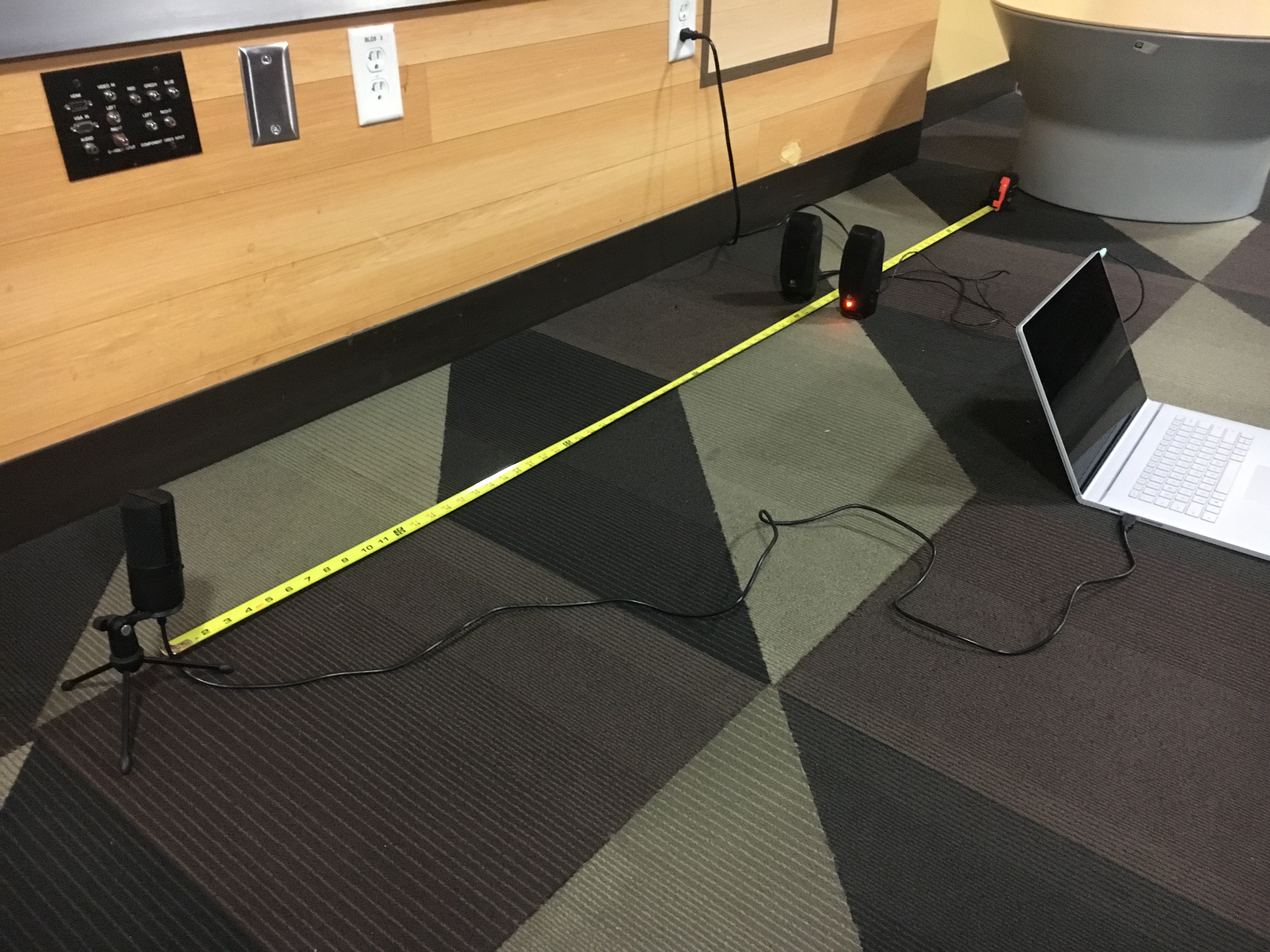}
            \caption[\scriptsize Dining Hall]%
            {{\scriptsize Dining Hall}}    
            \label{fig:mean and std of net24}
        \end{subfigure}
        \hfill
        \begin{subfigure}[b]{0.244\textwidth}   
            \centering 
            \includegraphics[width=\textwidth]{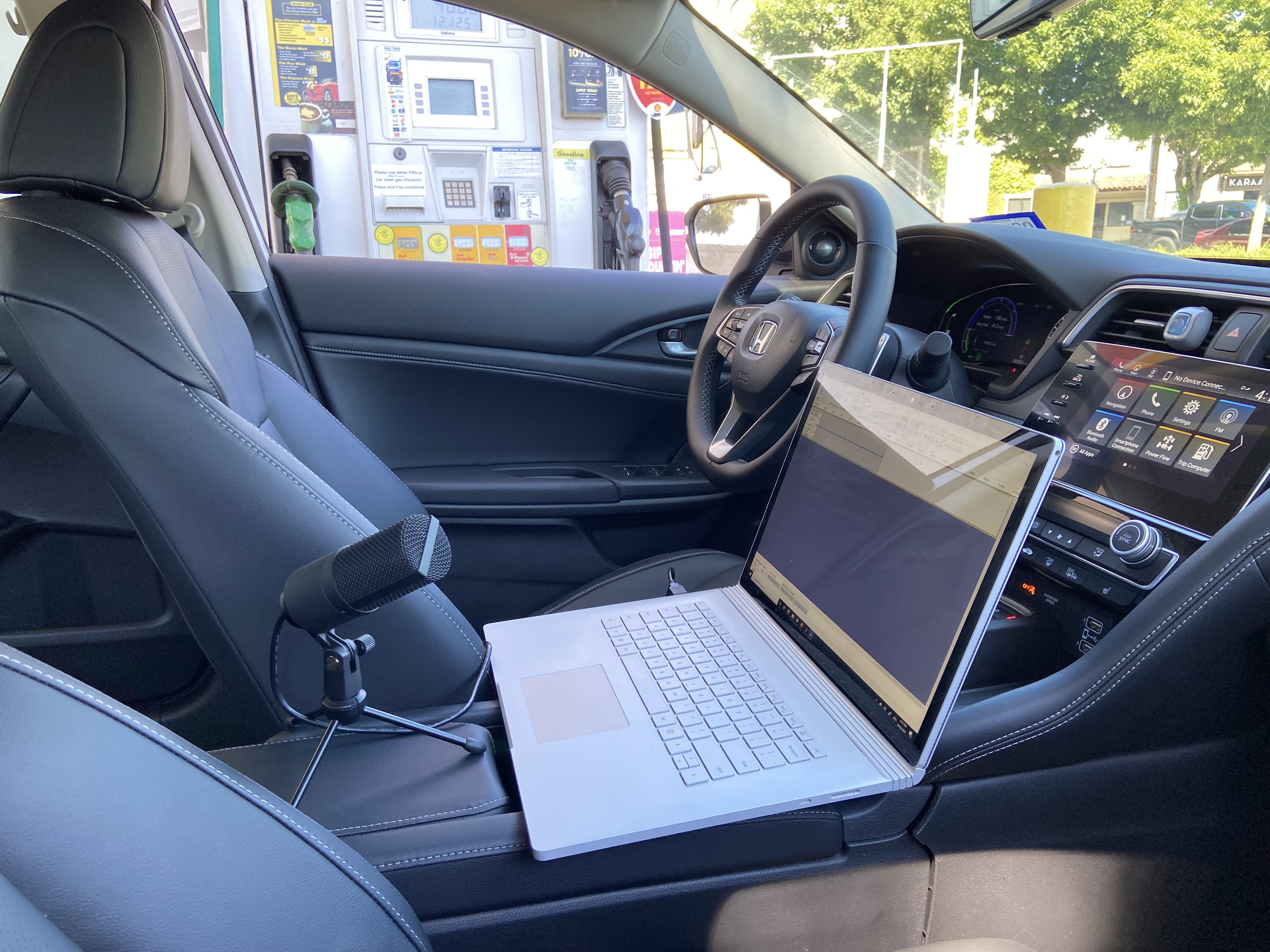}
            \caption[]%
            {{\scriptsize Gas Station (Car).}}    
            \label{fig:mean and std of net34}
        \end{subfigure}
        \hfill
        \begin{subfigure}[b]{0.244\textwidth}   
            \centering 
            \includegraphics[width=\textwidth]{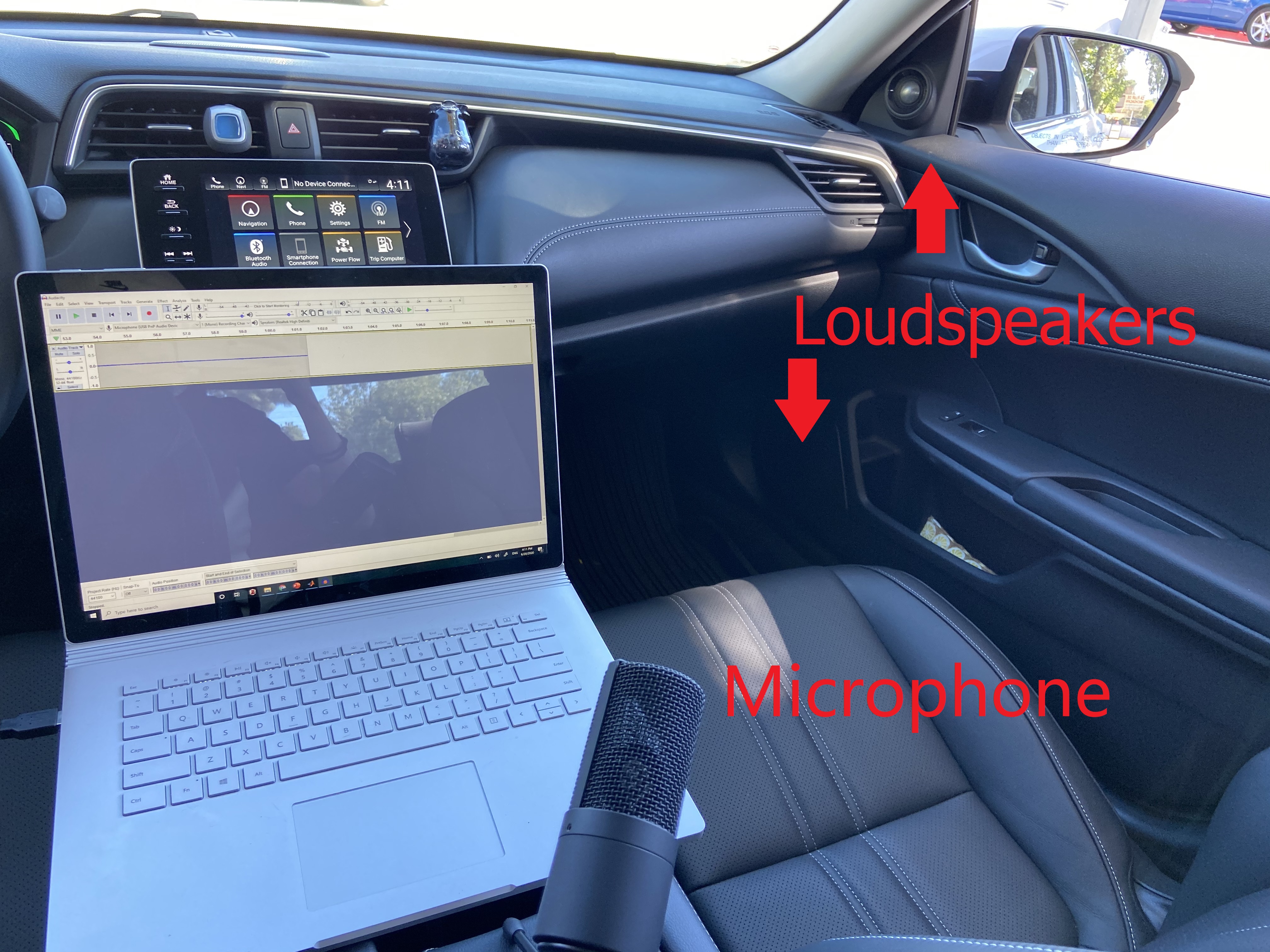}
            \caption[]%
            {{\scriptsize Setup in the Car.}}    
            \label{fig:mean and std of net44}
        \end{subfigure}
        \caption[]
        {\small Experimental setup to evaluate \tool\ in different environments.}
        \label{fig:realdevice}
    \end{figure}

\tool\ is implemented as a prototype in MATLAB and evaluated in three different environments: (i) a conference room (approx. 5.5 m by 4 m) selected as a reference for a semi-public indoor environment, (ii) a campus dining area (approx. 25m by 50m) chosen to mimic a noisy indoor public area, and (iii) inside a car parked at a gas station to simulate the use case of voice-driven payments at gas stations~\cite{gasstationpay}. For (i) and (ii), a FIFINE K669B microphone is used to record speech, and a Logitech S120 Speaker System is used as playback device to simulate users. The evaluation is performed at different distances (0.5-4 m) between the microphone and playback device. For (iii), the microphone is identical but the car's in-built speakers are used instead of the Logitech speakers. Also, the evaluation is performed at two different distances 0.5 and 0.7 m.

\tool\ is evaluated both on speech data captured from live participants in ReMASC dataset~\cite{gong2019remasc} and synthesized speech from Amazon Polly. A total of ten different speaker profiles are used (four of which are synthesized). For each profile, three different speech samples are used.
Microsoft Azure's speech and speaker recognition services~\cite{ms_speaker} are used for speech-to-text and speaker recognition. Text-independent speaker verification service~\cite{ms_verification} is used for speaker verification experiments. Each speaker's profile is first enrolled using a speech sample longer than 20 seconds. Verification is performed using the speaker's speech samples with and without acoustic nonce, and yields one of the following outputs: acceptance with ``high'', ``normal'', or ``low'' confidence, or a rejection.

\subsection{Performance}
\label{subsec:eval_performance}

\begin{table}[t]
\centering
\scriptsize
\caption{Optimal parameters for achieving reliability and imperceptibility at a distance of 0.5 m between the microphone and user in different environments: (a) conference room, (b) dining hall, and (c) in a car parked at a gas station.}
\label{tab:parameters}
\begin{tabular}{lllll}
  \toprule[1.5pt]
 Location  & \parbox[t]{1.5cm}{SPL (dB-A)}&\parbox[t]{1.8cm}{Frequency \\($\pm$ 200 Hz)} & \parbox[t]{1.8cm}{Amplitude \\ (normalized)} & \parbox[t]{1.1cm}{Bitrate \\  (bits/sec)}\\
  \midrule
 \parbox[t]{2.5cm}{Conference Room} & 41 & 4000 Hz & 0.52& 35\\
 Dining Hall & 58 & 5200 Hz & 1 & 25 \\
 Gas Station (Car) & 47 & 4800 Hz & 0.61 & 33 \\
    \bottomrule
\end{tabular}
\end{table}

\smallskip \noindent \textbf{Optimal Parameters.}
First, we estimate optimal parameters for achieving reliability and imperceptibility simultaneously in a given environment using Algorithm~\ref{alg:de}. The frequency step size is set to 200 Hz and $\theta$ is set as 0.1 to ensure that the generated nonce is below the average human speech SPL of 60 db. It is observed that lower amplitudes and frequencies and higher bitrates are comparatively optimal for acoustic nonce embedding in low noise environments, e.g., conference room (see Table~\ref{tab:parameters}). Relatively higher amplitudes and frequencies, and lower bitrates are more optimal in environments with higher noise, such as the dining hall. This is because the background noise in such environments typically vests as a low amplitude low frequency noise.

\smallskip \noindent \textbf{Computational Overhead.} 
Algorithm~\ref{alg:de} iteratively optimizes the parameter set $\mathcal{P}$ for the given environment. For conference room, a single iteration is sufficient. However, for dining hall and gas station environments, a few iterations ($<10$) are required and the computation takes a few seconds. This does not cause a delay at the time of user interaction because it overlaps with the time a user typically takes to set up a \vcs. The acoustic nonce embedding and decoding process occurs while user is interacting with \vcs. The computational overhead is negligible because the nonce size is small, and the playback device is used only when the user is speaking.

\smallskip \noindent \textbf{Reliability}
Next, we measure the reliability of embedding and retrieving acoustic nonce in the three environments. For this, we compute the average BER over 5 independent trials at different distances between the \vcs input device (microphone) and user. Figure~\ref{fig:real_dis} shows that in all three environments, the average BER upto a distance of 1.5 m is 0\%. Note, however that BER increases significantly beyond 1.5m in dining hall because the environment is relatively noisy. 

\begin{table}[t]
\centering
\scriptsize
\caption{Acoustic nonce recovery bit error rate (BER) and speech re-use detection performance (FAR and FRR) at different distances in the three environments. }
\label{tab:detection}
\begin{tabular}{l|l|l|c|l|c|l|l|c}
  \toprule[1.5pt]
 Location & Dis.  & BER/FAR/FRR & Location & Dis.  & BER/FAR/FRR & Location & Dis.  & BER/FAR/FRR\\
  \midrule
\parbox[t]{1.35cm}{Conference\\Room} & 0.5m  & 0\%/0\%/0\% &\parbox[t]{0.9cm}{ Dining\\Hall} & 0.5m & 0.2\%/0\%/0.5\% & \parbox[t]{0.9cm}{Gas\\Station \\ (Car)} & 0.5m & 0\%/0\%/0\%\\
 & 1m & 0\%/0\%/0\% &  & 1m & 1.2\%/0\%/ 1\%& & 0.7m & 0.5\%0\%0.5\%\\
 & 4m & 1.2\%/0\%/0.5\% & & 2m & 75\%/0\% /78\% &&\\
  
    \bottomrule
\end{tabular}

\end{table}

\smallskip \noindent \textbf{Speech Re-use Detection.}
In this experiment, we measure the efficacy of \tool\ in detecting speech re-use (see Table~\ref{tab:detection}). The input speech is considered re-used if (i) the current nonce is not decoded correctly, and (ii) if any trace of an obsolete nonce is detected. FRR indicates the proportion of falsely rejected samples because of incorrect nonce decoding or because the detected nonce is obsolete. FAR indicates how many re-used speech samples with obsolete nonces are falsely accepted. \tool\ achieves 0.5\%  FRR  at  0\%  FAR  for  speech  re-use prevention upto a distance of 4 meters in the three environments.

\subsection{Robustness}

\smallskip \noindent \textbf{Impact on Speaker and Speech Recognition.} The goal of this experiment is to assess the impact of acoustic nonce embedding and decoding on the performance of a commercial speaker and speech recognition system (Microsoft Azure). The results show that there is limited impact on the performance of the speaker recognition system for the conference room and dining hall environments upto a distance of 1.5 m (see Appendix Figure~\ref{fig:identest}). The embedded nonce marginally increases the average word error rate (WER) of the speech recognition system to 4.4\%. 
The WER is close to the reported average WER of 5.1\% for the same system in the NIST 2000 Switchboard test set~\cite{xiong2018microsoft}. Manually reviewing the results indicates that most errors are related to incorrect suffix ``-ing''. A few errors are due to small word omissions such as ``an'' and ``a''. Relatively more errors occur in the dining hall environment due to the background noise. 

\begin{table*}[htbp]
    \caption{Performance of Microsoft Azure speaker recognition system (similarity score 0 - 1) and speech recognition system (WER) after nonce removal using two different FHSS schemes. Without nonce removal, the respective average similarity score and WER of the two systems are 0.897 and 5.1\%~\cite{xiong2018microsoft}. Lossy compression and equalization nonce removal techniques failed to remove the nonce and the speech samples were rejected by \tool.}

\centering
\scriptsize
\label{table:speaker_iden}
\begin{tabular}{c|c|c|c|c}
\toprule[1.5pt]
\multirow{2}{*}{Removal Technique}  &  \multicolumn{2}{c}{\parbox[t]{4.4cm}{ 4-5kHz, 2 Hop., Avg. of 30 samples}} & \multicolumn{2}{|c}{\parbox[t]{4.5cm}{ 1-8kHz, 10 Hop., Avg. of 30 samples}} \\
\cmidrule{2-5}
& \parbox{2cm}{\centering Speaker\\ Verification (Score, 0-1) }& \parbox{2.2cm}{\centering Speech \\ Recognition (WER \%)} &\parbox{2cm}{\centering Speaker\\ Verification (Score, 0-1) }& \parbox{2.2cm}{\centering Speech \\ Recognition (WER \%)}\\
    \midrule
Resampling    & 0.233 & 29.73\%& 0.170& 91.83\% \\
Amplitude Compression   & 0.217 & 34.31\% & 0.184 & 63.4\% \\
Filtering   & 0.623& 17.12\% & 0.319& 62.21\%\\
Additive Noise     & 0.318 & 53.13\% & 0.192 & 96.34\%\\
Lossy Compression    & Reject & Reject&   Reject & Reject\\
Equalization     & Reject & Reject&  Reject& Reject\\

    \bottomrule
\end{tabular}

\end{table*}

\smallskip \noindent \textbf{Acoustic Nonce Removal.}
An adversary may attempt to remove the embedded acoustic nonce and obtain a ``clean'' speech sample to conduct re-use attacks. To test the robustness of \tool\ in this adversarial setting, we use 6 common audio watermark removal techniques~\cite{wu2005efficiently} (see Appendix~\ref{appendix:nonce_removal} for details). Each technique is iteratively used until the acoustic nonce cannot be detected in a speech sample. The removal attempts are repeated for all speech samples in the dataset mentioned above. Table~\ref{table:speaker_iden} shows that nonce removal techniques reduced the average recognition performance of the commercial speaker recognition system by 79.52\%, and increased the average word error rate of the speech recognition system by 33.57\%.

\subsection{Human Perception}

\subsubsection{Empirical Measurement.} 
In this experiment, we empirically measure the change in SPL due to the embedded nonce. The distance between \vcs output and the recording device is 1 m in dining hall and conference room, and 0.7 m in the car. The results (see Appendix Figure~\ref{fig:dbdifference}) show that the average SPL difference in each environment is less than 1.4\%. Non-parametric Wilcoxon Signed Rank test~\cite{wilcoxon1970critical} (at 95\% confidence level) is used to measure the statistical significance of the difference in SPL before and after acoustic nonce embedding. The results indicate that the differences are statistically insignificant.

\begin{table*}[b]
\centering
\scriptsize
\caption{Perception of speech samples. Overall (\%) is the proportion of participants that label speech samples as imperceptible, non-disruptive, or disruptive. Sample (\%) is the proportion of speech samples (with nonce) that are determined to be imperceptible, non-disruptive, or disruptive by the majority of participants.}
\vspace{-0mm}
\label{tab:userstudy_overview}
\begin{tabular}{l|c|c|c}
  \toprule[1.5pt]
 Location & \parbox[t]{2.9cm}{Imperceptible \\ (Overall/Sample\%)}  & \parbox[t]{2.9cm}{Non-disruptive \\ (Overall/Sample\%)}  &\parbox[t]{3.1cm}{Disruptive, Nonce-caused \\ (Overall/Sample\%)}\\
  \midrule
 Conference Room & 34.87\%/50\% & 50\%/50\% & 1.31\%/0\% \\
 Dining Hall & 32.64\%/25\% & 41.67\%/75\% & 7.64\%/0\%\\
 Gas Station (Car) & 32.14\%/25\% & 42.86\%/75\% & 8.57\%/0\% \\
    \bottomrule
\end{tabular}
\end{table*}

\smallskip\noindent\textbf{User Study.}
Next, we study the impact of \tool\ on \vcs usability. For this, 120 participants (with prior experience using a \vcs) from various demographics (age groups, gender shown in Appendix Table~\ref{tab:group_userstudy}) are recruited on Amazon Mechanical Turk. Each participant is presented with a survey that takes approximately 10 minutes to complete. Post completion of the survey, a participant is paid 4 USD as incentive. The survey questions are designed to ensure bias is not introduced in the results (no leading questions).

Each participant is asked to enter their demographic information. The participant is then instructed to play a test sample and adjust the playback volume to a comfortable level. Following this, the participant is presented with a survey with five speech samples and follow up questions. Included in the five samples is one sample with only background noise (and no speech) to check if the participant is attentive. The participants are asked to provide speech transcriptions for all samples and the provided transcriptions are used to check the participants' attentiveness. Eleven participants failed to pass this test and hence were excluded.

The test samples are uniformly selected at random from a dataset of 24 speech recordings, 12 recordings each with and without embedded nonce from the three environments. The average length of the speech samples is 6.4 seconds. To ensure the user study is unbiased, all samples presented to a participant are from the same speaker and pertain to the same voice command. After listening to a speech sample, a participant is asked to rate the noise level of the speech sample on a scale of 0 - 10 (10 being the highest noise level). The participant is also instructed to report any noticeable noise source that affects their experience, and to specify the characteristic of the noise that resulted in this disruption (e.g., human conversation, train or car noise). The answer is used to ascertain the source of noise that impacted the usability (e.g., the embedded nonce).

\smallskip \noindent \textbf{Results.} Table~\ref{tab:userstudy_overview} reports the average human perception ratings. The acoustic nonce does not degrade overall user experience for 94.16\% of speech samples, on average, in the three environments. Since this does not adequately measure the participant's perception of nonce embedding for each sample, per speech sample perception is also reported. No speech sample with embedded nonce is perceived to be disruptive by the majority of participants.

\begin{table}[h]
\centering
\scriptsize
\caption{Average perceived noise (on a scale of 0-10) in speech samples with and without acoustic nonce collected in the three environments.}
\label{tab:noise_userstudy}
\begin{tabular}{lcccc}
  \toprule[1.5pt]
 Location &  \parbox[t]{1.7cm}{w/o Nonce} & \parbox[t]{1.7cm}{w/ Nonce} & t-test, $\alpha=0.05$ & Statistically Significant \\
  \midrule
 Conference Room & 3.8 & 4.8 & $t =4.2022, p =0.0001489$ & Yes\\
 Dining Hall & 5.7 & 6.1 & $t=1.4595, p =0.1524$ & No\\
 Gas Station (Car) & 5 & 5.6 &  $t=3.0697, p=0.00389$ & Yes\\
    \bottomrule
\end{tabular}

\end{table}

\smallskip \noindent \textbf{Perceived Noise Level.}
Table~\ref{tab:noise_userstudy} shows that the speech samples containing the acoustic nonce are perceived to have a higher noise level. The dependent samples paired t-test (at 95\% confidence) indicates a statistically significant difference (between samples with and without nonce) in the conference room and car. However, this difference is statistically insignificant in the dining hall due to higher background noise level compared to the conference room and car. Despite perceiving higher noise level, the majority of participants do not perceive any speech sample to be disruptive (see Table~\ref{tab:userstudy_overview}).
\section{Summary and Future Work}
In this work, we present a security overlay called \tool\ that can be integrated with any voice-driven service to prevent speech re-use. \tool\ reliably embeds a dynamic acoustic nonce that is non-disruptive to a \vcs user, and detects the embedded nonce from the recorded user speech. \tool\ can be combined with other defense mechanisms to further enhance \vcs security. As an example, it can be used alongside a dynamic challenge-response method where in each interaction, the user is expected to respond to a fresh challenge (e.g., repeat a phrase). The acoustic nonce can be embedded in each challenge, and can be detected using \tool. Experiments conducted in three different environments show that \tool\ can be used in operational \vcs scenarios with minimal overhead. In future, we plan to conduct in-depth investigation of methods for generation and encoding of acoustic nonce, e.g., using variations of music or other forms of sound along with different encoding schemes. We also plan to conduct large-scale live subject testing to further evaluate reliability and imperceptibility.

\bibliographystyle{plain}
\bibliography{main}
\begin{appendix}
\section{Appendix: Distance between adversary and \vcs speaker.}

\begin{figure}
    \centering
    \begin{subfigure}[b]{0.4\textwidth}
    \includegraphics[width=\textwidth]{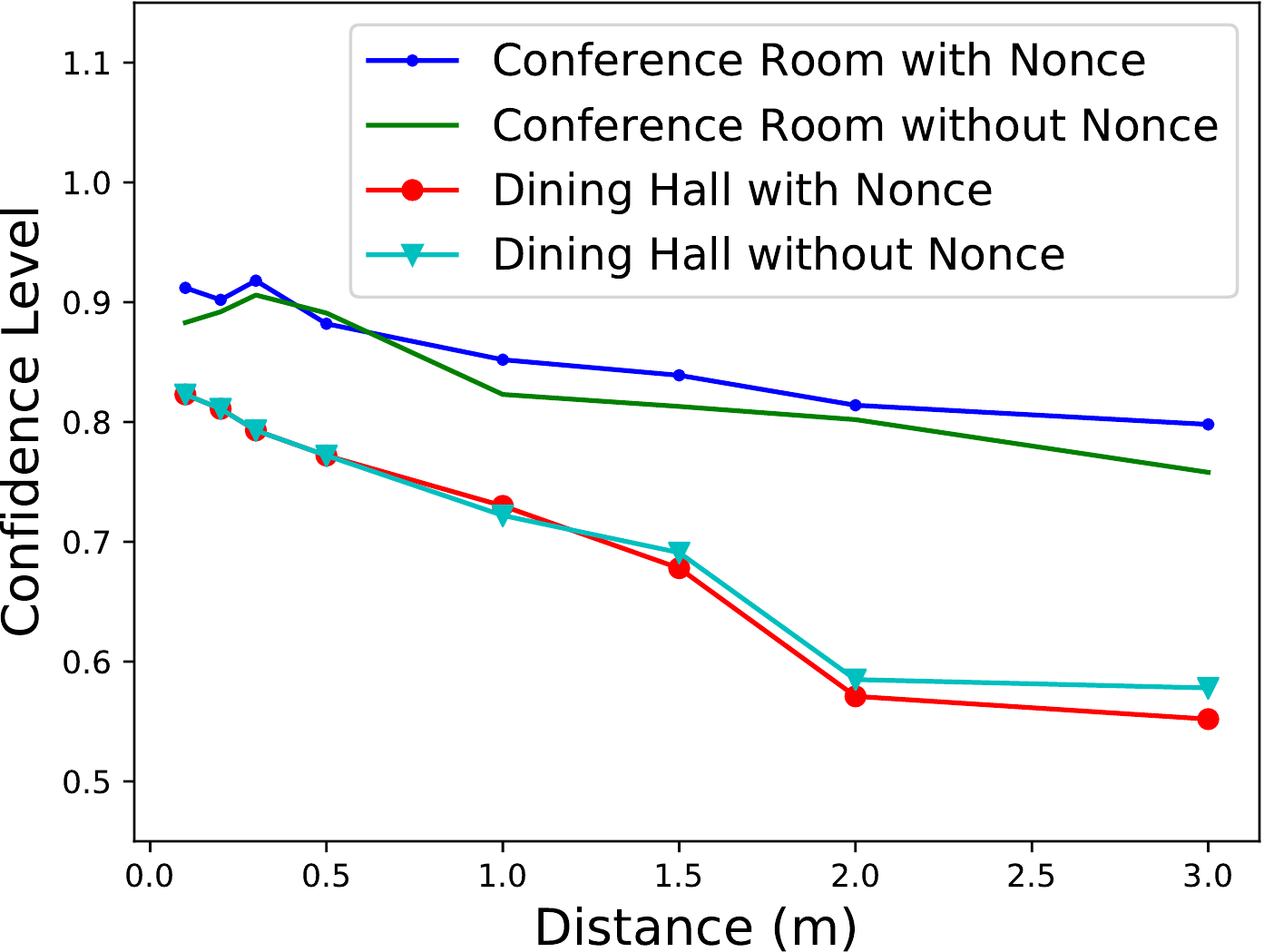}
    \caption{}
    \label{fig:identest}
    \end{subfigure}
    \hfill
    \begin{subfigure}[b]{0.4\textwidth}  
        \includegraphics[width=\textwidth]{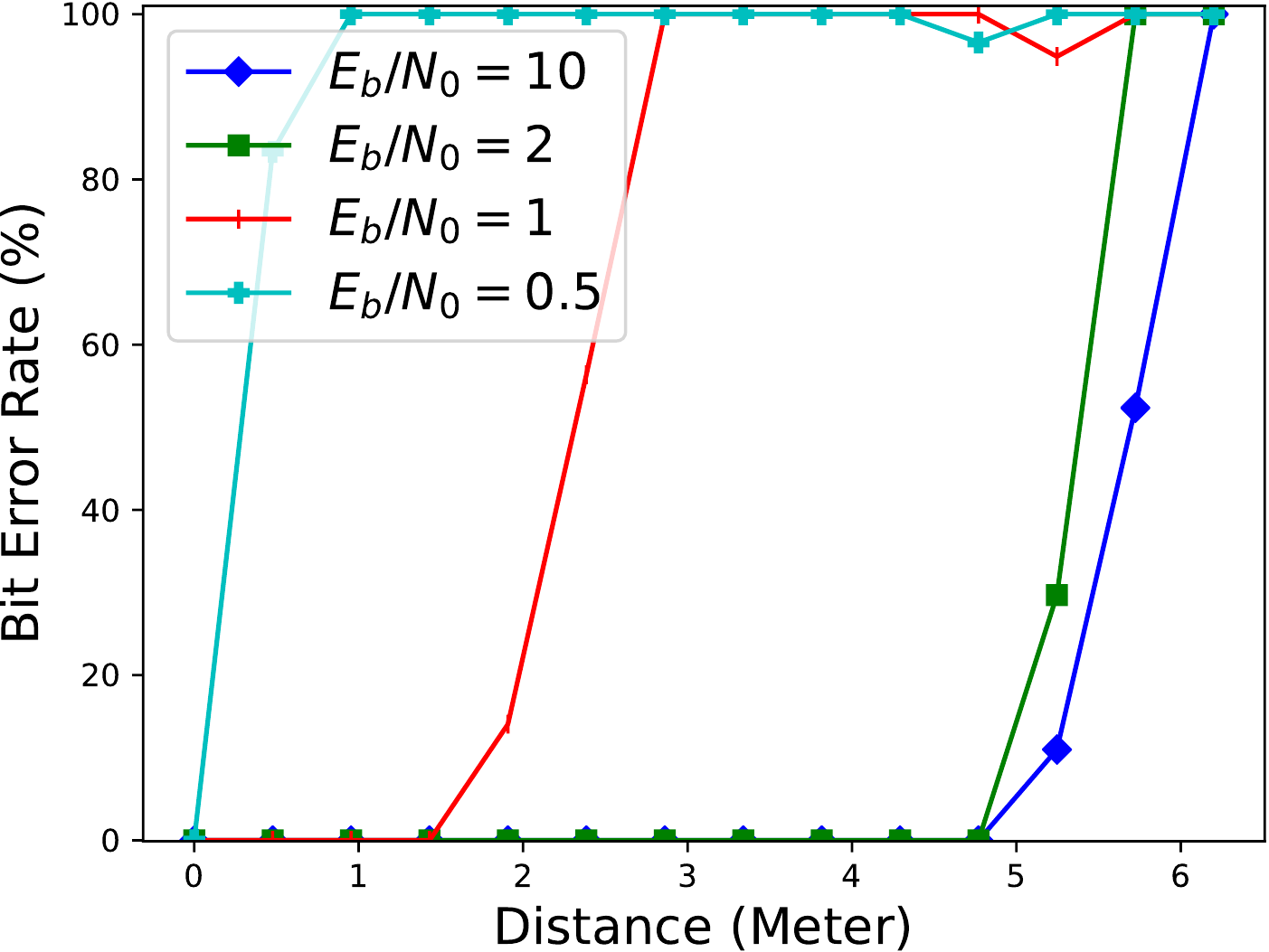}
        \caption{}
        \label{fig:BER_dist}
    \end{subfigure}
    \caption{\small (a) Impact of the proposed security overlay on prediction confidence of Microsoft Azure Speaker Recognition system in three different environments. (b) Impact of distance between the \vcs speaker and an adversary's recording device at the time of capture on speech re-use.  }
\end{figure}

To launch a speech re-use attack, an adversary first records a user's interaction with \vcs. In this experiment, we measure how effectively the acoustic nonce is preserved with varying distance between the \vcs speaker and adversary's recording device (microphone) at the time of capture. Figure~\ref{fig:BER_dist} shows that this distance significantly impacts BER. The computed BER is 0\% upto a distance of 4 m when the signal to noise ratio is low. If the captured speech is being re-used, assuming a lossless audio input to \vcs, the entire nonce can be recovered and used for determining the particular user interaction session the speech was previously used in. 

\begin{figure}[h]
\centering
  \includegraphics[width=0.4\linewidth]{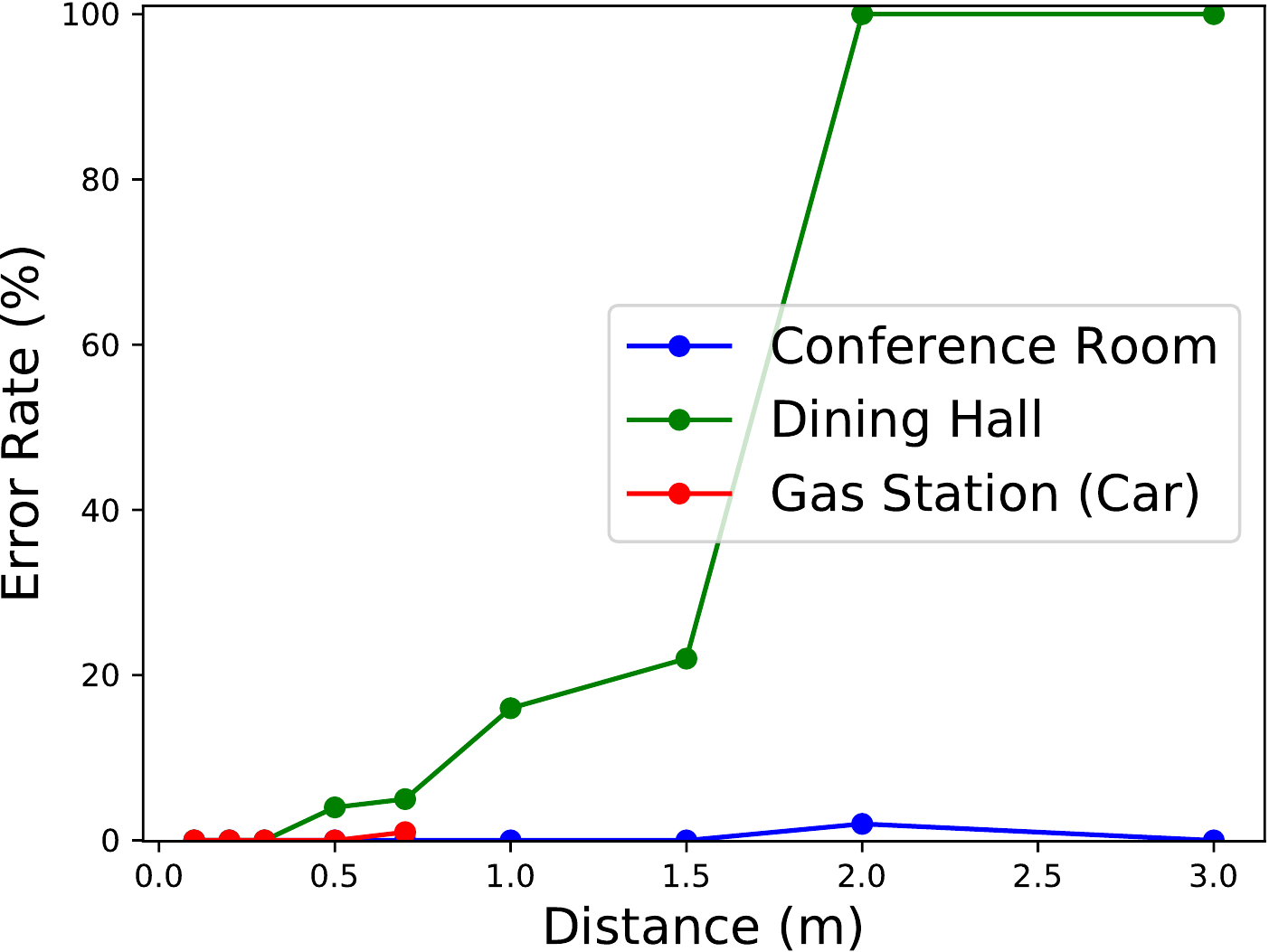}
  \caption{Variation of bit error rate (\%) with different distances between microphone and user (simulated using a playback device) in different environments: (a) conference room, (b) dining hall, and (c) in a car parked at a gas station. Given a fixed size car in (c), results can only be computed upto a distance of 0.7 m.}
  \label{fig:real_dis}
\end{figure}

\begin{figure}[t]
    \centering 
    \includegraphics[width=0.4\textwidth]{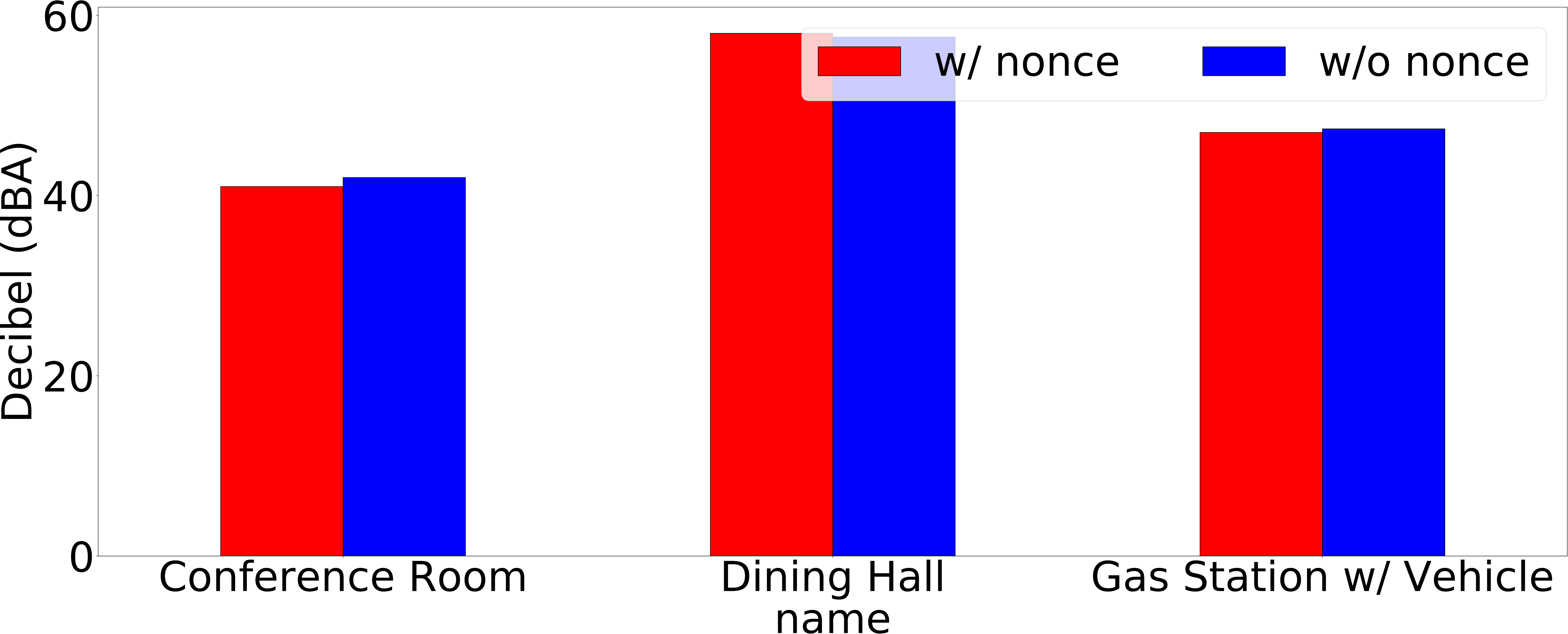}
   \caption{Loudness of speech with and without embedded nonce (average value of 5 measurements).}

    \label{fig:dbdifference}
\end{figure}

\section{Appendix: Nonce Removal Techniques}
\label{appendix:nonce_removal}

\begin{itemize}[noitemsep,leftmargin=*]
    \item Resampling. Samples the audio at a different frequency (e.g. 44.1 KHz) to remove sampling-dependent watermark.
    \item Amplitude Compression. Alters the amplitude of recorded audio to bypass amplitude-related watermark detection.
    \item Filtering. Uses a high pass or low pass filter to remove a specific range of frequency in the audio. For example, separate audible signals from inaudible frequencies (e.g. ultrasound).  
    \item Lossy Compression. Leverages data encoding methods that uses inexact approximations and partial data discarding to compress the audio. For example, MP3~\cite{mp3} or MPEG-1 Audio Layer III reduce audio file size by taking advantage of a perceptual limitation of human hearing.
    \item Equalization. Modifies the frequency response of an audio system using linear filters. Frequency response is the measure of the audio output in comparison to the input as a function of frequency. For example, a 5kHz wave input may have a 4kHz output response.
    \item Additive Noise. Adds noise to decrease SNR (signal-to-noise ratio) so that watermark cannot be extracted successfully.
\end{itemize}

\section{Appendix: User Study Demographics}
The demographics for the 120 participants in the user study are shown in Figure~\ref{tab:group_userstudy}. There are more female participants than male participants, and 56.7\% of the participants are between age 25 to 35. 

\begin{table}[h]
\centering
\small
\vspace{-5mm}
\caption{User study participant demographic information.}
\vspace{-0mm}
\label{tab:group_userstudy}
\begin{tabular}{l|ccc|cccc}
  \toprule[1.5pt]
  \multirow{2}{*}{Category} & \multicolumn{3}{c}{Gender Groups} & \multicolumn{4}{c}{Age Groups} \\
  \cmidrule{2-8}
 & Male&Female&Others &  18-25, &25-35,  &35-45, & 45-55\\
 \midrule
 Number & 65 &53 &2  & 42 & 68  & 41 & 11\\

    \bottomrule
\end{tabular}
\end{table}

\end{appendix}

\end{document}